\documentclass{article}

\usepackage{arxiv}

\usepackage[utf8]{inputenc} 
\usepackage[T1]{fontenc}    
\usepackage{hyperref}       
\usepackage{url}
\usepackage{booktabs}       
\usepackage{amsfonts}       
\usepackage{nicefrac}       
\usepackage{microtype}      
\usepackage{lipsum}
\usepackage{graphicx}
\graphicspath{ {./images/} }
\usepackage{fancyhdr}
\usepackage{tabu}
\usepackage{array}
\usepackage{wasysym, pifont}
\usepackage{tabularx,longtable}
\usepackage{makecell}
\usepackage{booktabs, geometry}
\usepackage{float}
\usepackage{rotating}
\usepackage{multicol}
\usepackage[title]{appendix}
\usepackage[super]{nth}

\newcolumntype{?}{!{\vrule width 1pt}}
\usepackage{multirow}
\usepackage{blindtext}
\usepackage{graphicx}
\renewcommand{\arraystretch}{1.5}
\usepackage{xcolor,colortbl}
\definecolor{greeny}{HTML}{53bca3}
\definecolor{bluey}{HTML}{BF1158}
\definecolor{yellowy}{HTML}{FFA000}
\definecolor{grayy}{HTML}{BCBCBC}
\definecolor{Mycolor1}{HTML}{F2DFD5}
\usepackage{amssymb}
\usepackage{booktabs}
\graphicspath{ {./images/} }

\title{Exploring the Relationships between Privacy by Design Schemes and Privacy Laws: A Comparative Analysis}

\author{
 Atheer Aljeraisy \\
  School of Computer Science and Informatics\\
  Cardiff University, UK \\
  \texttt{aljeraisya@cardiff.ac.uk} \\
   \And
  Masoud Barati \\
Department of Computer and Information Science\\
Cardiff University, UK \\
\texttt{BaratiM@cardiff.ac.uk} \\
  \And
    Omer Rana \\
  School of Computer Science and Informatics\\
  Cardiff University, UK \\
  \texttt{RanaOF@cardiff.ac.uk} \\
  \And
  Charith Perera \\
  School of Computer Science and Informatics\\
  Cardiff University, UK \\
  \texttt{PereraC@cardiff.ac.uk} \\
}
\usepackage{subcaption}

\begin{document}
\maketitle
\begin{abstract}
Internet of Things (IoT) applications have the potential to derive sensitive information about individuals. Therefore, developers must exercise due diligence to make sure that data are managed according to the privacy regulations and data protection laws. However, doing so can be a difficult and challenging task. Recent research has revealed that developers typically face difficulties when complying with regulations. One key reason is that, at times, regulations are vague, and could be challenging to extract and enact such legal requirements.  In our research paper, we have conducted a systematic analysis of the data protection laws that are used across different continents, namely: (i) General Data Protection Regulations (GDPR), (ii) the Personal Information Protection and Electronic Documents Act (PIPEDA), (iii) the California Consumer Privacy Act (CCPA), (iv) Australian Privacy Principles (APPs), and (v) New Zealand’s Privacy Act 1993. In this technical report, we presented the detailed results of the conducted framework analysis method to attain a comprehensive view of different data protection laws and highlighted the disparities, in order to assist developers in adhering to the regulations across different regions, along with creating a Combined Privacy Law Framework (CPLF). After that, we gave an overview of various Privacy by Design (PbD) schemes developed previously by different researchers. Then, the key principles and individuals’ rights of the CPLF were mapped with the privacy principles, strategies, guidelines, and patterns of the Privacy by Design (PbD) schemes in order to investigate the gaps in existing schemes. 
\end{abstract}



\maketitle

\newpage
\section{Report Structure} \label{PaperStructure}
The technical report contains five sections and is structured as follows: Section \ref{PaperStructure} presents paper structure. Section \ref{Appendix:Key Definitions} presents the key definitions of the selected data protection laws. Section \ref{Appendix:Result} shows the detailed results of the key principles and individuals’ rights that resulted from the analysis process (i.e., Combined Privacy Law Framework (CPLF)). Section \ref{Appendix:Result} shows the detailed results of the Combined Privacy Laws Framework (CPLF) using framework analysis. Section \ref{Appendix:Privacy by Design Schemes} presents various Privacy by Design (PbD) schemes (e.g., privacy principles, strategies, guidelines, and patterns) that developed previously by different researchers. In Section \ref{Appendix:Mapping between the Combined Privacy Laws Framework and Privacy by Design Schemes}, the correlation of the principles and rights of the CPLF and the principles, strategies, guidelines, and patterns of the PbD schemes are demonstrated.
\section{Key Definitions}\label{Appendix:Key Definitions}

This section articulates various definitions of different terms that are used in the regulations of the data protection laws of General Data Protection Regulations (GDPR) \cite{GDPRMain}, the Personal Information Protection and Electronic Documents Act (PIPEDA) \cite{PIPEDAPrinciples, OfficeoPrivacyCommissionerCanada}, California Consumer Privacy Act (CCPA) \cite{BillTextAB-375Privacy}, Australian Privacy Principles (APPs) \cite{Appsquickreference, RightsResponsibilitiesOAIC}, and the New Zealand Privacy Act (1993) \cite{Commissioneroftheprivacyprinciples,YourRightToKnow}.

\subsection{Personal Data}

\begin{itemize}

\item{\textbf{GDPR:}}
‘Personal Data’ means ‘any information relating to an identified or identifiable natural person (‘data subject’); an identifiable natural person is one who can be identified, directly or indirectly, in particular by reference to an identifier such as a name, an identification number, location data, an online identifier or to one or more factors specific to the physical, physiological, genetic, mental, economic, cultural or social identity of that natural person’;

\item{\textbf{PIPEDA:}}
‘Personal Data’ is defined under PIPEDA as ‘Personal Information’ that includes ‘any factual or subjective information, recorded or not, about an identifiable individual. This includes information in any form, such as:
age, name, ID numbers, income, ethnic origin, or blood type;
opinions, evaluations, comments, social status, or disciplinary actions; and
employee files, credit records, loan records, medical records, existence of a dispute between a consumer and a merchant, intentions (for example, to acquire goods or services, or change jobs)’.

\item{\textbf{CCPA:}}
 ‘Personal Data’ is defined under CCPA as ‘Personal information’ that means ‘information that identifies, relates to, describes, is capable of being associated with, or could reasonably be linked, directly or indirectly, with a particular consumer or household’.

\item{\textbf{APPs:}}
‘‘Personal Data’ is defined under the APPS as ‘Personal Information’ which means ‘any information or an opinion about an identified individual, or an individual who is reasonably identifiable:
\begin{itemize}
    \item whether the information or opinion is true or not; and 
	\item whether the information or opinion is recorded in a material form or not’.
\end{itemize}

\item{\textbf{New Zealand Privacy Act 1993:}}
‘Personal Data’ is defined under the Privacy Act as ‘Personal Information’ which means ‘information about an identifiable individual’.
\end{itemize}


\subsection{Data Subject}

\begin{itemize}

 \item{\textbf{GDPR:}}
‘Data Subject’ means ‘an identifiable natural person who can be identified directly or indirectly’.

 \item {\textbf{PIPEDA:}}
‘Data Subject’ is defined under PIPEDA as ‘Individuals’.

 \item {\textbf{CCPA:}}
‘Data Subject’ is defined under CCPA as “Consumer” which means ‘a natural person who is a California resident, as defined in Section 17014 of Title 18 of the California Code of Regulations, as that section read on September 1, 2017, however identified, including by any unique identifier’.

 \item {\textbf{APPs:}}
‘Data Subject’ is defined under APPS as ‘Individuals’, which means ‘a natural person’.

 \item {\textbf{New Zealand Privacy Act 1993:}}
‘Data Subject’ is defined under the Privacy Act 1993 as ‘Individuals’, which means ‘a natural person’.

\end{itemize}


\subsection{Processing}

\begin{itemize}

 \item{\textbf{GDPR:}}
‘processing’ means ‘any operation or set of operations which is performed on personal data or on sets of personal data, whether or not by automated means, such as collection, recording, organisation, structuring, storage, adaptation or alteration, retrieval, consultation, use, disclosure by transmission, dissemination or otherwise making available, alignment or combination, restriction, erasure or destruction’;

 \item{\textbf{PIPEDA:}}
‘Processing’ under PIPEDA would include the use, collection, disclosure, alteration, storage or destruction of personal information.

 \item{\textbf{CCPA:}}
‘Processing’ is defined under CCPA ‘as any operation or set of operations that are performed on personal data or on sets of personal data, whether or not by automated means. These operations would include collection and selling that are defined as:

\begin{itemize}

\item{‘Collects,’ ‘collected,’ or ‘collection’ means buying, renting, gathering, obtaining, receiving, or accessing any personal information pertaining to a consumer by any means. This includes receiving information from the consumer, either actively or passively, or by observing the consumer’s behavior.}

\item{‘Sell, ‘selling,’ ‘sale,’ or ‘sold,’ means selling, renting, releasing, disclosing, disseminating, making available, transferring, or otherwise communicating orally, in writing, or by electronic or other means, a consumer’s personal information by the business to another business or a third party for monetary or other valuable consideration’.}

\end{itemize}

\item{\textbf{APPs:}}
‘Processing’ under APPS would include ‘collection’, ‘deal’, ‘use’, and ‘disclose’.

\begin{itemize}

\item{‘Collection’ is ‘applied broadly, and includes gathering, acquiring or obtaining personal information from any source and by any means’.}
\item{‘Use’ is not defined in the Privacy Act. ‘An APP entity ‘uses’ information where it handles or undertakes an activity with the information, within the entity’s effective control’.}
\item{‘Disclose’ is not defined in the Privacy Act. ‘An APP entity ‘discloses’ personal information where it makes it accessible to others outside the entity and releases the subsequent handling of the information from its effective control’.}

\end{itemize}

\item{\textbf{New Zealand Privacy Act 1993:}}
‘Processing’ under The Privacy Act 1993 would include ‘collect’, ‘use’, and ‘disclose’.

\end{itemize}


\subsection{Controller}

\begin{itemize}

\item{\textbf{GDPR:}}
‘Controller’ means ‘the natural or legal person, public authority, agency or other body which, alone or jointly with others, determines the purposes and means of the processing of personal data; where the purposes and means of such processing are determined by Union or Member State law, the controller or the specific criteria for its nomination may be provided for by Union or Member State law’;

\item{\textbf{PIPEDA:}}
‘Controller’ is referred under PIPEDA to ‘organizations’ which means ‘private sector organizations that are not federally regulated. It does not apply to organizations that do not engage in commercial, for-profit activities’.

\item{\textbf{CCPA:}}
 ‘Controller’ is referred under CCPA to ‘business’, which means ‘a sole proprietorship, partnership, limited liability company, corporation, association, or other legal entity that is organized or operated for the profit or financial benefit of its shareholders or other owners, that does business in the State of California , and that satisfies one or more of the following thresholds:
 
 \begin{enumerate}
    
\item{Has annual gross revenues in excess of twenty-five million dollars (\$25,000,000), as adjusted pursuant to paragraph (5) of subdivision (a) of Section 1798.185.}
\item{Alone or in combination, annually buys, receives for the business’ commercial purposes, sells, or shares for commercial purposes, alone or in combination, the personal information of 50,000 or more consumers, households, or devices.}
\item{Derives 50 percent or more of its annual revenues from selling consumers’ personal information’.}

 \end{enumerate}
 
\item{\textbf{APPs:}}
‘Controller’ is referred under APPS to ‘App entity’, which defined to be an ‘agency’ or ‘organization’ which means ‘an individual (including a sole trader), a body corporate, a partnership, any other unincorporated association, or a trust, unless it is a small business operator, registered political party, State or Territory authority or a prescribed instrumentality of a State’.

\item{\textbf{New Zealand Privacy Act 1993:}}
‘Controller’ is referred under The Privacy Act 1993 to ‘agency’, which means ‘any person or body of persons, whether corporate or unincorporated, and whether in the public sector or the private sector; and, for the avoidance of doubt, includes a department’.

\end{itemize}


\subsection{Processor}

\begin{itemize}

\item {\textbf{GDPR:}}
'processor' means ‘a natural or legal person, public authority, agency or other body which processes personal data on behalf of the controller’;

\item {\textbf{PIPEDA:}}
‘Processor’ is referred under PIPEDA to ‘organizations’.

\item {\textbf{CCPA:}}
 ‘Processor’ is referred under PIPEDA to ‘business’.

\item {\textbf{APPs:}}
‘Processor’ is referred under APPS to ‘App entity’, which defined to be an “agency” or “organization”.

\item{\textbf{New Zealand Privacy Act 1993:}}
‘Processor’ is referred under the Privacy Act 1993 to an ‘agency’.

\end{itemize}


\section{Result}\label{Appendix:Result}
This section presents the result of a high-level comparison of various data protection laws, which are: General Data Protection Regulations (GDPR) \cite{GDPRMain}, the Personal Information Protection and Electronic Documents Act (PIPEDA) \cite{PIPEDAPrinciples, OfficeoPrivacyCommissionerCanada}, California Consumer Privacy Act (CCPA) \cite{BillTextAB-375Privacy}, Australian Privacy Principles (APPs) \cite{Appsquickreference, RightsResponsibilitiesOAIC}, and the New Zealand Privacy Act (1993) \cite{Commissioneroftheprivacyprinciples,YourRightToKnow}, through a framework analysis method according to our research paper to attain a comprehensive view of various data protection laws and highlight the disparities. This will assist developers in adhering to the regulations across different regions.. The analysis of the Combined Privacy Laws Framework (CPLF) \footnote{Combined Privacy Laws Framework refers to the selected data protection laws for this study.} is based on the principles and rights.

\vspace{5em}

\begingroup
\footnotesize
\begin{table}[htbp!] 
\begin{minipage}{\columnwidth}
\caption{Transparency Principle}

        \begin{tabular}{c?l|l}
        \multicolumn{3}{ p{\textwidth}} {\textbf{Definition:} \textit{The organisation must provide detailed information on its policies and procedures concerning the management of personal information, and this must be readily available to the public.}} \\ \toprule[0.32ex]
        \multirow{2}{50pt}{GDPR} & Article 5 and Recital 39 & \textit{lawfulness, fairness and transparency} \\ \cline{2-3}
        & \multicolumn {2}{ p{0.85\textwidth}} {\scriptsize{‘Personal data shall be processed lawfully, fairly and in a transparent manner in relation to the data subject’, where ‘(39) The principle of transparency requires that any information and communication relating to the processing of those personal data be easily accessible and easy to understand, and that clear and plain language be used’. \newline
        (39) The principle of transparency requires that any information and communication relating to the processing of those personal data be easily accessible and easy to understand, and that clear and plain language be used.}}  \\
        \toprule
        \multirow{2}{50pt} {PIPEDA} & Principle 8 & \textit{Openness} \\ \cline{2-3}
        & \multicolumn {2}{ p{0.85\textwidth}} {\scriptsize{‘An organization must make detailed information about its policies and practices relating to the management of personal information publicly and readily available’.}}  \\
        \toprule
        \multirow{2}{50pt} {CCPA} & - & \textit{-} \\ \cline{2-3}
        & \multicolumn {2}{ p{0.85\textwidth}} {\scriptsize{Not mentioned explicitly as a principle or right, but it says in the introduction that: ‘Our desire for privacy controls and transparency in data practices is heightened’.}}  \\
        \toprule
        \multirow{2}{50pt} {APPS} & App1 & \textit{Open and transparent management of personal information } \\ \cline{2-3}
        & \multicolumn {2}{ p{0.85\textwidth}} {\scriptsize{‘Ensures that APP entities manage personal information in an open and transparent way. This includes having a clearly expressed and up to date APP privacy policy’.}}  \\
        \toprule
        \multirow{2}{50pt} {New Zealand} & Principle 3 & \textit{Collection of information from subject} \\ \cline{2-3}
        & \multicolumn {2}{ p{0.85\textwidth}} {\scriptsize{‘Agencies should be open when collecting information.... The best way to do this is usually with a clear privacy statement’.}} \\
        \toprule [0.32ex]
    \end{tabular}
    \label{tab:Transparency Principle}
\end{minipage}

\end{table}
\endgroup


\begingroup
\footnotesize
\begin{table}[htbp!]
\caption{Purpose Limitation Principle}
\begin{minipage}{\columnwidth}

        \begin{tabular}{c?l|l}
        \multicolumn{3}{ p{\textwidth}} {\textbf{Definition:} \textit{The purposes that the personal information is being collected and is used for must be shared by the organization, either prior to or at the time it is collected.}} \\ 
        \toprule[0.32ex]
        \multirow{2}{50pt}{GDPR} & Article 5 & \textit{Purpose Limitation} \\ \cline{2-3}
        & \multicolumn {2}{ p{0.85\textwidth}} {\scriptsize{‘Personal data shall be collected for specified, explicit and legitimate purposes and not further processed in a manner that is incompatible with those purposes; further processing for archiving purposes in the public interest, scientific or historical research purposes or statistical purposes shall, in accordance with Article 89(1), not be considered to be incompatible with the initial purposes’;}}  \\
        \toprule
        \multirow{2}{50pt} {PIPEDA} & Principle 2 & \textit{Identifying Purposes} \\ \cline{2-3}
        & \multicolumn {2}{ p{0.85\textwidth}} {\scriptsize{‘The purposes for which the personal information is being collected must be identified by the organization before or at the time of collection’.}}  \\
        \toprule
        \multirow{2}{50pt} {CCPA} & 1798.100. (b) & \textit{-} \\ \cline{2-3}
        & \multicolumn {2}{ p{0.85\textwidth}} {\scriptsize{‘A business shall not collect additional categories of personal information or use personal information collected for additional purposes without providing the consumer with notice consistent with this section’.}}  \\
        \toprule
        \multirow{2}{50pt} {APPS} & App 6 & \textit{Use or disclosure of personal information} \\ \cline{2-3}
        & \multicolumn {2}{ p{0.85\textwidth}} {\scriptsize{6.12 ‘The purpose for which an APP entity collects personal information is known as the ‘primary purpose’ of collection. This is the specific function or activity for which the entity collects the personal information’.}}  \\
        \toprule
        \multirow{2}{50pt} {New Zealand} & Principle 1 & \textit{Purpose for collection of personal information} \\ \cline{2-3}
        & \multicolumn {2}{ p{0.85\textwidth}} {\scriptsize{‘Agencies need to carefully consider the purpose for which they collect personal information’.}}  \\
        \toprule [0.32ex]
    \end{tabular}
    \label{tab:Purpose Limitation Principle}

\end{minipage}

\vspace{5em}

\footnotesize
\caption{Limiting Use and Disclosure Principle}

\begin{minipage}{\columnwidth}
        \begin{tabular}{c?l|l}
        \multicolumn{3}{ p{\textwidth}} {\textbf{Definition:} \textit{An organisation may only use or disclose personal information for the purposes for which it was collected.}} \\ 
        \toprule[0.32ex]
        \multirow{2}{50pt}{GDPR} & Article 5 & \textit{Purpose limitation} \\ \cline{2-3}
        & \multicolumn {2}{ p{0.85\textwidth}} {\scriptsize{‘Personal data shall be collected for specified, explicit and legitimate purposes and not further processed in a manner that is incompatible with those purposes; further processing for archiving purposes in the public interest, scientific or historical research purposes or statistical purposes shall, in accordance with Article 89(1), not be considered to be incompatible with the initial purposes’}}  \\
        \toprule
        \multirow{2}{50pt} {PIPEDA} & Principle 5 & \textit{Limiting Use, Disclosure, and Retention} \\ \cline{2-3}
        & \multicolumn {2}{ p{0.85\textwidth}} {\scriptsize{‘Unless the individual consents otherwise or it is required by law, personal information can only be used or disclosed for the purposes for which it was collected’.}}  \\
        \toprule
        \multirow{2}{50pt} {CCPA} & 1798.100. (b) and 1798.120.(b) & \textit{-} \\ \cline{2-3}
        & \multicolumn {2}{ p{0.85\textwidth}} {\scriptsize{1798.100. (b) ‘A business shall not …….  use personal information collected for additional purposes without providing the consumer with notice consistent with this section’.
1798.120.(b) ‘A business that sells consumers’ personal information to third parties shall provide notice to consumers, pursuant to subdivision (a) of Section 1798.135, that this information may be sold and that consumers have the “right to opt-out” of the sale of their personal information’.}}  \\
        \toprule
        \multirow{2}{50pt} {APPs} & App 6 & \textit{Use or disclosure of personal information} \\ \cline{2-3}
        & \multicolumn {2}{ p{0.85\textwidth}} {\scriptsize{‘An APP entity can only use or disclose personal information for a purpose for which it was collected (known as the ‘primary purpose’), or for a secondary purpose if an exception applies’.}}  \\
        \toprule
        \multirow{2}{50pt} {New Zealand} & Principle10 - 11 & \textit{10 + 11: Limits on use and disclosure of personal information} \\ \cline{2-3}
        & \multicolumn {2}{ p{0.85\textwidth}} {\scriptsize{P10: ‘Agencies must use personal information for the same purpose for which they collected that information’.
P11: ‘Agencies can only disclose personal information in limited circumstances. One example is where another law requires them to disclose the information’.}}  \\
        \toprule [0.32ex]
    \end{tabular}
    \label{tab:Limiting Use and Disclosure Principle}
\end{minipage}
\end{table}
\endgroup


\begingroup
\footnotesize
\begin{table}[htbp!]
\caption{Data Minimisation Principle}
\begin{minipage}{\columnwidth}
        \begin{tabular}{c?l|l}
        \multicolumn{3}{ p{\textwidth}} {\textbf{Definition:} \textit{Where personal data is required, it must be relevant, adequate and limited only to what is needed for the purpose stated.}} \\ 
        \toprule[0.32ex]
        \multirow{2}{50pt}{GDPR} & Article 5 & \textit{Data Minimisation} \\ \cline{2-3}
        & \multicolumn {2}{ p{0.85\textwidth}} {\scriptsize{‘Personal data shall be adequate, relevant and limited to what is necessary in relation to the purposes for which they are processed’.}}  \\
        \toprule
        \multirow{2}{50pt} {PIPEDA} & Principle 4 & \textit{Limiting Collection} \\ \cline{2-3}
        & \multicolumn {2}{ p{0.85\textwidth}} {\scriptsize{‘The collection of personal information must be limited to that which is needed for the purposes identified by the organization. Information must be collected by fair and lawful means’.}}  \\
        \toprule
        \multirow{2}{50pt} {CCPA} & 1798.100. (b) & \textit{-} \\ \cline{2-3}
        & \multicolumn {2}{ p{0.85\textwidth}} {\scriptsize{‘A business shall not collect additional categories of personal information … for additional purposes without providing the consumer with notice consistent with this section’.}}  \\
        \toprule
        \multirow{2}{50pt} {APPS} & App 3 & \textit{Collection of solicited personal information} \\ \cline{2-3}
        & \multicolumn {2}{ p{0.85\textwidth}} {\scriptsize{‘An agency may collect personal information that is ‘directly related to’ one or more of the agency’s functions or activities (APP 3.1). To be ‘directly related to’, a clear and direct connection must exist between the personal information being collected and an agency function or activity (APP 3.16)’.}}  \\
        \toprule
        \multirow{2}{50pt} {New Zealand} & Principle 1 & \textit{Purpose for collection of personal information} \\ \cline{2-3}
        & \multicolumn {2}{ p{0.85\textwidth}} {\scriptsize{‘Personal information must only be collected when: the collection is for a lawful purpose, connected with what the agency does, and it’s necessary to collect the information for that purpose’.}}  \\
        \toprule [0.32ex]
    \end{tabular}
    
    \label{tab:Data Minimisation Principle}
\end{minipage}

\vspace{3em}


\footnotesize
\caption{Consent Principle}
\begin{minipage}{\columnwidth}
        \begin{tabular}{c?l|l}
        \multicolumn{3}{ p{\textwidth}} {\textbf{Definition:} \textit{The individual’s consent is required for the collection, use, and/or disclosure of their personal information.}} \\ 
        \toprule[0.32ex]
        \multirow{2}{50pt}{GDPR} & Recital 11 - Article 7 – Article 8  & \textit{Conditions for consent - Conditions applicable to child's consent ...} \\ \cline{2-3}
        & \multicolumn {2}{ p{0.85\textwidth}} {\scriptsize{‘consent of the data subject means any freely given, specific, informed and unambiguous indication of the data subject's wishes by which he or she, by a statement or by a clear affirmative action, signifies agreement to the processing of personal data relating to him or her’;
 ‘Where processing is based on consent, the controller shall be able to demonstrate that the data subject has consented to processing of his or her personal data’.
 ‘Where the child is below the age of 16 years, such processing shall be lawful only if and to the extent that consent is given or authorised by the holder of parental responsibility over the child’.}}  \\
        \toprule
        \multirow{2}{50pt} {PIPEDA} & Principle 3 & \textit{Consent} \\ \cline{2-3}
        & \multicolumn {2}{ p{0.85\textwidth}} {\scriptsize{‘Organizations are generally required to obtain meaningful consent for the collection, use and disclosure of personal information’.
‘To make consent meaningful, people must understand what they are consenting to’.
 ‘Consent and children: Obtain consent from a parent or guardian for any individual unable to provide meaningful consent themselves (the OPC takes the position that, in all but exceptional’}}  \\
        \toprule
        \multirow{2}{50pt} {CCPA} & 1798.120. (c) - 1798.120. (d) & \textit{-} \\ \cline{2-3}
        & \multicolumn {2}{ p{0.85\textwidth}} {\scriptsize{1798.120. (c) ‘A business that has received direction from a consumer not to sell the consumer’s personal information or, in the case of a minor consumer’s personal information has not received consent to sell the minor consumer’s personal information shall be prohibited, pursuant to paragraph (4) of subdivision (a) of Section 1798.135, from selling the consumer’s personal information after its receipt of the consumer’s direction, unless the consumer subsequently provides express authorization for the sale of the consumer’s personal information’.
1798.120. (d) ‘Notwithstanding subdivision (a), a business shall not sell the personal information of consumers if the business has actual knowledge that the consumer is less than 16 years of age, unless the consumer, in the case of consumers between 13 and 16 years of age, or the consumer’s parent or guardian, in the case of consumers who are less than 13 years of age, has affirmatively authorized the sale of the consumer’s personal information’.
}}  \\
        \toprule
        \multirow{2}{50pt} {APPS} & Chapter B & \textit{Key concepts} \\ \cline{2-3}
        & \multicolumn {2}{ p{0.85\textwidth}} {\scriptsize{B.35 ‘Consent means ‘express consent or implied consent’ (s 6(1)). The four key elements of consent are: the individual is adequately informed before giving consent, the individual gives consent voluntarily, the consent is current and specific, and the individual has the capacity to understand and communicate their consent’.
B.48 ‘An APP entity should generally seek consent from an individual for collection and proposed uses and disclosures of personal information at the time the information is collected’.  
B.57 ‘As a general principle, an individual under the age of 18 has capacity to consent when they have sufficient understanding and maturity to understand what is being proposed. In some circumstances, it may be appropriate for a parent or guardian to consent on behalf of a young person’.
}}  \\
        \toprule
        \multirow{2}{50pt} {New Zealand} & Principle 3 & \textit{Collection of information from subject} \\ \cline{2-3}
        & \multicolumn {2}{ p{0.85\textwidth}} {\scriptsize{N/A
New Zealand’s Privacy Act does not require the agency to obtain an individual’s consent, however, ‘agencies must take reasonable steps to ensure that the person they are collecting information from is aware’.}}  \\
        \toprule [0.32ex]
    \end{tabular}
    
   \label{tab:Consent Principle}
\end{minipage}
\end{table}
\endgroup

\begingroup
\footnotesize
\begin{table}[htbp!]
\caption{Lawfulness of processing Principle}
\begin{minipage}{\columnwidth}
        \begin{tabular}{c?l|l}
        \multicolumn{3}{ p{\textwidth}} {\textbf{Definition:} \textit{Personal information must be processed by fair and lawful means.}} \\ 
        \toprule[0.32ex]
        \multirow{2}{50pt}{GDPR} & Article 6 & \textit{Lawfulness of Processing} \\ \cline{2-3}
        & \multicolumn {2}{ p{0.85\textwidth}} {\scriptsize{‘Processing shall be lawful only if and to the extent that at least one of the following applies: (a) the data subject has given consent to the processing of his or her personal data for one or more specific purposes; (b) processing is necessary for the performance of a contract to which the data subject is party or in order to take steps at the request of the data subject prior to entering into a contract; (c) processing is necessary for compliance with a legal obligation to which the controller is subject; (d) processing is necessary in order to protect the vital interests of the data subject or of another natural person; (e) processing is necessary for the performance of a task carried out in the public interest or in the exercise of official authority vested in the controller; (f) processing is necessary for the purposes of the legitimate interests pursued by the controller or by a third party, except where such interests are overridden by the interests or fundamental rights and freedoms of the data subject which require protection of personal data, in particular where the data subject is a child’.}}  \\
        \toprule
        \multirow{2}{50pt} {PIPEDA} & Principle 4 & \textit{Limiting Collection} \\ \cline{2-3}
        & \multicolumn {2}{ p{0.85\textwidth}} {\scriptsize{‘Information must be collected by fair and lawful means. This requirement is intended to prevent organizations from collecting information by misleading or deceiving about the purpose’.}}  \\
        \toprule
        \multirow{2}{50pt} {CCPA} & - & \textit{-} \\ \cline{2-3}
        & \multicolumn {2}{ p{0.85\textwidth}} {\scriptsize{Not mentioned explicitly, but it can be inferred from 1798.105. (d): ‘A business or a service provider shall not be required to comply with a consumer’s request to delete the consumer’s personal information if it is necessary for the business or service provider to maintain the consumer’s personal information in order to............: (9) Otherwise use the consumer’s personal information, internally, in a lawful manner that is compatible with the context in which the consumer provided the information’.}}  \\
        \toprule
        \multirow{2}{50pt} {APPS} & App 3 & \textit{Collection of solicited personal information} \\ \cline{2-3}
        & \multicolumn {2}{ p{0.85\textwidth}} {\scriptsize{‘An APP entity must collect personal information ‘only by lawful and fair means’ (APP 3.5). This requirement applies to all APP entities’.}}  \\
        \toprule
        \multirow{2}{50pt} {New Zealand} & Principle 4 & \textit{Manner of collection of personal information} \\ \cline{2-3}
        & \multicolumn {2}{ p{0.85\textwidth}} {\scriptsize{‘Personal information must not be collected by unlawful means or by means that are unfair or unreasonably intrusive in the circumstances’.}}  \\
        \toprule [0.32ex]
    \end{tabular}
   \label{tab:Lawfulness of processing Principle}
\end{minipage}

\vspace{5em}


\footnotesize
\caption{Accuracy Principle}
\begin{minipage}{\columnwidth}
        \begin{tabular}{c?l|l}
        \multicolumn{3}{ p{\textwidth}} {\textbf{Definition:} \textit{Attempts must be made to ensure that personal information is as accurate, complete and up-to-date as possible to ensure it satisfies the purposes for which it will be used.}} \\ 
        \toprule[0.32ex]
        \multirow{2}{50pt}{GDPR} & Article 5 & \textit{Accuracy} \\ \cline{2-3}
        & \multicolumn {2}{ p{0.85\textwidth}} {\scriptsize{Attempts must be made to ensure that personal information is as accurate, complete and up-to-date as possible to ensure it satisfies the purposes for which it will be used.}}  \\
        \toprule
        \multirow{2}{50pt} {PIPEDA} & Principle 6 & \textit{Accuracy} \\ \cline{2-3}
        & \multicolumn {2}{ p{0.85\textwidth}} {\scriptsize{‘Personal information must be as accurate, complete, and up-to-date as possible in order to properly satisfy the purposes for which it is to be used’.}}  \\
        \toprule
        \multirow{2}{50pt} {CCPA} & - & \textit{-} \\ \cline{2-3}
        & \multicolumn {2}{ p{0.85\textwidth}} {\scriptsize{N/A}}  \\
        \toprule
        \multirow{2}{50pt} {APPS} & App 10 & \textit{Quality of personal information} \\ \cline{2-3}
        & \multicolumn {2}{ p{0.85\textwidth}} {\scriptsize{‘An APP entity must take reasonable steps to ensure the personal information it collects is accurate, up to date and complete. An entity must also take reasonable steps to ensure the personal information it uses or discloses is accurate, up to date, complete and relevant, having regard to the purpose of the use or disclosure’.}}  \\
        \toprule
        \multirow{2}{50pt} {New Zealand} & Principle 8 & \textit{Accuracy, etc, of personal information to be checked before use} \\ \cline{2-3}
        & \multicolumn {2}{ p{0.85\textwidth}} {\scriptsize{‘Before it uses or discloses personal information, an agency must take reasonable steps to check that information is accurate, complete, relevant, up to date and not misleading’.}}  \\
        \toprule [0.32ex]
    \end{tabular}
    
  \label{tab:Accuracy Principle}
\end{minipage}
\end{table}
\endgroup


\begingroup
\footnotesize
\begin{table}[htbp!]
\caption{Storage Limitation Principle}
\begin{minipage}{\columnwidth}
        \begin{tabular}{c?l|l}
        \multicolumn{3}{ p{\textwidth}} {\textbf{Definition:} \textit{Personal information must not be stored for longer than necessary once it has met the purposes for which it was collected.}} \\ 
        \toprule[0.32ex]
        \multirow{2}{50pt}{GDPR} & Article 5 & \textit{Storage limitation} \\ \cline{2-3}
        & \multicolumn {2}{ p{0.85\textwidth}} {\scriptsize{‘Personal data shall be kept in a form which permits identification of data subjects for no longer than is necessary for the purposes for which the personal data are processed; personal data may be stored for longer periods insofar as the personal data will be processed solely for archiving purposes in the public interest, scientific or historical research purposes or statistical purposes in accordance with Article 89(1) subject to implementation of the appropriate technical and organisational measures required by this Regulation in order to safeguard the rights and freedoms of the data subject’.}}  \\
        \toprule
        \multirow{2}{50pt} {PIPEDA} & Principle 5 & \textit{Limiting Use, Disclosure, and Retention } \\ \cline{2-3}
        & \multicolumn {2}{ p{0.85\textwidth}} {\scriptsize{‘Personal information must only be kept as long as required to serve those purposes. Destroy, erase or anonymise any personal information that your organization no longer needs’.}}  \\
        \toprule
        \multirow{2}{50pt} {CCPA} & 1798.140. & \textit{-} \\ \cline{2-3}
        & \multicolumn {2}{ p{0.85\textwidth}} {\scriptsize{Not mentioned as a requirement for the businesses, however, Service provider and third party are ‘[prohibited] from retaining, using, or disclosing the personal information for any purpose other than for the specific purpose of performing the services specified in the contract for the business’.}}  \\
        \toprule
        \multirow{2}{50pt} {APPS} & App 11 & \textit{Security of personal information } \\ \cline{2-3}
        & \multicolumn {2}{ p{0.85\textwidth}} {\scriptsize{11.3 ‘An APP entity must take reasonable steps to destroy or de-identify the personal information it holds once the personal information is no longer needed for any purpose for which the personal information may be used or disclosed under the APPs’.}}  \\
        \toprule
        \multirow{2}{50pt} {New Zealand} & Principle 9 & \textit{Agency not to keep personal information for longer than necessary } \\ \cline{2-3}
        & \multicolumn {2}{ p{0.85\textwidth}} {\scriptsize{‘An agency that holds personal information must not keep that information for longer than is necessary for the purposes for which the information may be lawfully used’.}}  \\
        \toprule [0.32ex]
    \end{tabular}
    
    \label{tab:Storage Limitation Principle}
\end{minipage}

\vspace{5em}


\footnotesize
\caption{Security Principle}
\begin{minipage}{\columnwidth}
        \begin{tabular}{c?l|l}
        \multicolumn{3}{ p{\textwidth}} {\textbf{Definition:} \textit{Safeguarding procedures must be in place to protect personal information and prevent its loss, misuse or disclosure.}} \\ 
        \toprule[0.32ex]
        \multirow{2}{50pt}{GDPR} & Article 5 – A 33 & \textit{integrity and confidentiality - Notification of a personal data breach ...} \\ \cline{2-3}
        & \multicolumn {2}{ p{0.85\textwidth}} {\scriptsize{‘Personal data shall be processed in a manner that ensures appropriate security of the personal data, including protection against unauthorised or unlawful processing and against accidental loss, destruction or damage, using appropriate technical or organisational measures.
‘In the case of a personal data breach the controller shall without undue delay and, where feasible, not later than 72 hours after having become aware of it notify the personal data breach to the supervisory authority’.}}  \\
        \toprule
        \multirow{2}{50pt} {PIPEDA} & Principle 7 & \textit{Safeguards - What you need to know about mandatory reporting of breaches ...} \\ \cline{2-3}
        & \multicolumn {2}{ p{0.85\textwidth}} {\scriptsize{‘Personal information must be protected by appropriate security relative to the sensitivity of the information’.
‘Organizations subject to the (PIPEDA) will be required to: report to the Privacy Commissioner of Canada breaches of security safeguards involving personal information that pose a real risk of significant harm to individuals, notify affected individuals about those breaches, and keep records of all breaches’ \cite{reportingofbreachesofsecurity} }}  \\
        \toprule
        \multirow{2}{50pt} {CCPA} & - & \textit{-} \\ \cline{2-3}
        & \multicolumn {2}{ p{0.85\textwidth}} {\scriptsize{Not mentioned explicitly, but it can be inferred from 1798.140. (h) (1) that business that uses deidentified information must have “implemented technical safeguards that prohibit reidentification of the consumer to whom the information may pertain.”
Existing law requires a business or person that suffers a breach of security of computerized data that includes personal information, as defined, to disclose that breach, as specified.}}  \\
        \toprule
        \multirow{2}{50pt} {APPS} & App 11 & \textit{Security of personal information - Australian entities and the EU GDPR} \\ \cline{2-3}
        & \multicolumn {2}{ p{0.85\textwidth}} {\scriptsize{‘An APP entity must take reasonable steps to protect personal information it holds from misuse, interference and loss, and from unauthorized access, modification or disclosure. An entity has obligations to destroy or de-identify personal information in certain circumstances’.
‘A notifiable data breaches scheme requires APP entities to provide a statement to the Commissioner notifying of an eligible data breach as soon as practicable after the entity becomes aware of the breach. It also requires entities to notify affected individuals as soon as practicable after preparing the statement for the Commissioner’\cite{AppsEU}. }}  \\
        \toprule
        \multirow{2}{50pt} {New Zealand} & Principle 5 & \textit{Storage and security of personal information - Do we have to report data breaches?} \\ \cline{2-3}
        & \multicolumn {2}{ p{0.85\textwidth}} {\scriptsize{‘It’s impossible to stop all mistakes. But agencies must ensure that there are reasonable safeguards in place to prevent loss, misuse or disclosure of personal information’.
‘Breach notification is voluntary but that is likely that will change with privacy law reform. The Government has indicated that a mandatory requirement to report data breaches is going to be part of the changes made in a new Privacy Act’ \cite{OfficePrivacyCommissioner}. }}  \\
        \toprule [0.32ex]
    \end{tabular}
    
    \label{tab:Security Principle}
\end{minipage}
\end{table}
\endgroup


\begingroup
\footnotesize
\begin{table}[htbp!]
\caption{Accountability Principle}
\begin{minipage}{\columnwidth}
        \begin{tabular}{c?l|l}
        \multicolumn{3}{ p{\textwidth}} {\textbf{Definition:} \textit{An organisation is responsible for complying with all data protection laws concerning the personal information that is under its control, and it must appoint someone to be accountable for complying with the law in this area.}} \\ 
        \toprule[0.32ex]
        \multirow{2}{50pt}{GDPR} & Article 24 - Article 37 & \textit{Accountability - Designation of the data protection officer} \\ \cline{2-3}
        & \multicolumn {2}{ p{0.85\textwidth}} {\scriptsize{ ‘The controller shall be responsible for, and be able to demonstrate compliance with, [law].’
‘The controller and the processor shall designate a data protection officer [in any of the specified cases]’.}}  \\
        \toprule
        \multirow{2}{50pt} {PIPEDA} & Principle 1 & \textit{Accountability} \\ \cline{2-3}
        & \multicolumn {2}{ p{0.85\textwidth}} {\scriptsize{‘An organization is responsible for personal information under its control. It must appoint someone to be accountable for its compliance with these fair information principles’.
‘The Privacy Officer should establish a program that demonstrates compliance by mapping the program to applicable legislation’.}}  \\
        \toprule
        \multirow{2}{50pt} {CCPA} & 1798.155. (a) & \textit{-} \\ \cline{2-3}
        & \multicolumn {2}{ p{0.85\textwidth}} {\scriptsize{ ‘Any business or third party may seek the opinion of the Attorney General for guidance on how to comply with the provisions of this title’.}}  \\
        \toprule
        \multirow{2}{50pt} {APPS} & App 1 & \textit{Open and transparent management of personal information} \\ \cline{2-3}
        & \multicolumn {2}{ p{0.85\textwidth}} {\scriptsize{1.4 ‘An APP entity must take reasonable steps to implement practices, procedures and systems that will ensure it complies with the APPs and any binding registered APP code, and is able to deal with related inquiries and complaints’.
10: ‘Privacy Officer: An agency must, at all times, have a designated Privacy Officer. An agency may have more than one Privacy Officer’.}}  \\
        \toprule
        \multirow{2}{50pt} {New Zealand} & - & \textit{23 Privacy Officer} \\ \cline{2-3}
        & \multicolumn {2}{ p{0.85\textwidth}} {\scriptsize{‘It shall be the responsibility of each agency to ensure that there are, within that agency, 1 or more individuals whose responsibilities include: 
(a) the encouragement of compliance, by the agency, with the information privacy principles.    (b) dealing with requests made to the agency pursuant to this Act.    (c) working with the Commissioner in relation to investigations conducted pursuant to Part 8 in relation to the agency.    (d) otherwise ensuring compliance by the agency with the provisions of this Act’.}}  \\
        \toprule [0.32ex]
    \end{tabular}
    \label{tab:Accountability Principle}
\end{minipage}

\vspace{5em}


\footnotesize
\caption{Anonymity and Pseudonymity Principle}
\begin{minipage}{\columnwidth}
        \begin{tabular}{c?l|l}
        \multicolumn{3}{ p{\textwidth}} {\textbf{Definition:} \textit{Individuals must be given the option of not identifying themselves, or the choice of using a pseudonym, in relation to certain matters.}} \\ 
        \toprule[0.32ex]
        \multirow{2}{50pt}{GDPR} & Article 32 & \textit{Security of processing} \\ \cline{2-3}
        & \multicolumn {2}{ p{0.85\textwidth}} {\scriptsize{Not mentioned for the purpose of dealing with an organisation, but mentioned under security of processing:
‘the controller and the processor shall implement appropriate technical and organisational measures to ensure a level of security appropriate to the risk, including inter alia as appropriate:
(a)	the pseudonymisation and encryption of personal data’
}}  \\
        \toprule
        \multirow{2}{50pt} {PIPEDA} & - & \textit{-} \\ \cline{2-3}
        & \multicolumn {2}{ p{0.85\textwidth}} {\scriptsize{Not mentioned for the purpose of dealing with an organisation, but mentioned in the case and principles of storage limitation:
‘Destroy, erase or anonymise any personal information that your organization no longer needs’.
}}  \\
        \toprule
        \multirow{2}{50pt} {CCPA} & 1798.140 & \textit{-} \\ \cline{2-3}
        & \multicolumn {2}{ p{0.85\textwidth}} {\scriptsize{It is not mentioned clearly, however, the CCPA referred the ‘Pseudonymise’ or ‘Pseudonymisation’ to ‘the processing of personal information in a manner that renders the personal information no longer attributable to a specific consumer without the use of additional information, provided that the additional information is kept separately and is subject to technical and organizational measures to ensure that the personal information is not attributed to an identified or identifiable consumer’.}}  \\
        \toprule
        \multirow{2}{50pt} {APPS} & App 2 & \textit{Anonymity and pseudonymity} \\ \cline{2-3}
        & \multicolumn {2}{ p{0.85\textwidth}} {\scriptsize{‘App2 Requires APP entities to give individuals the option of not identifying themselves, or of using a pseudonym’.}}  \\
        \toprule
        \multirow{2}{50pt} {New Zealand} & - & \textit{-} \\ \cline{2-3}
        & \multicolumn {2}{ p{0.85\textwidth}} {\scriptsize{N/A}}  \\
        \toprule [0.32ex]
    \end{tabular}
    \label{tab:Anonymity and Pseudonymity Principle}
\end{minipage}
\end{table}
\endgroup


\begingroup
\footnotesize
\begin{table}[htbp!]
\caption{Source Principle}
\begin{minipage}{\columnwidth}
        \begin{tabular}{c?l|l}
        \multicolumn{3}{ p{\textwidth}} {\textbf{Definition:} \textit{Unless there is an exception, personal information must be collected from the person the information is about.}} \\ 
        \toprule[0.32ex] 
        \multirow{2}{50pt}{GDPR} & Article 14 & \textit{Information to be provided where personal data have not been obtained ...} \\ \cline{2-3}
        & \multicolumn {2}{ p{0.85\textwidth}} {\scriptsize{2. ‘…, the controller shall provide the data subject with the following information necessary to ensure fair and transparent processing in respect of the data subject: … (f) from which source the personal data originate, and if applicable, whether it came from publicly accessible sources’;}}  \\
        \toprule
        \multirow{2}{50pt} {PIPEDA} & Principle 4 & \textit{Limiting Collection} \\ \cline{2-3}
        & \multicolumn {2}{ p{0.85\textwidth}} {\scriptsize{‘Upon request, an organization shall inform an individual whether or not the organization holds personal information about the individual. Organizations are encouraged to indicate the source of this information’.}}  \\
        \toprule
        \multirow{2}{50pt} {CCPA} & 1798.110. (a) & \textit{-} \\ \cline{2-3}
        & \multicolumn {2}{ p{0.85\textwidth}} {\scriptsize{‘A consumer shall have the right to request that a business that collects personal information about the consumer disclose to the consumer the following: 
(2) The categories of sources from which the personal information is collected’.
}}  \\
        \toprule
        \multirow{2}{50pt} {APPS} & App 3 & \textit{Collection of solicited personal information } \\ \cline{2-3}
        & \multicolumn {2}{ p{0.85\textwidth}} {\scriptsize{‘An organisation must, on request, provide its source for an individual’s personal information, unless it is impracticable or unreasonable to do so.
APP 3.6 provides that an APP entity ‘must collect personal information about an individual only from the individual’, unless exceptions apply’.
}}  \\
        \toprule
        \multirow{2}{50pt} {New Zealand} & Principle 2 & \textit{Source of personal information} \\ \cline{2-3}
        & \multicolumn {2}{ p{0.85\textwidth}} {\scriptsize{‘Personal information must usually be collected from the person the information is about. Collecting information from the person concerned means that people know what is going on and have some control over their information’.}}  \\
        \toprule [0.32ex]
    \end{tabular}
    \label{tab:Source Principle}
\end{minipage}

\vspace{5em}


\footnotesize
\caption{Cross-border Disclosure of Personal Information Principle}
\begin{minipage}{\columnwidth}
        \begin{tabular}{c?l|l}
        \multicolumn{3}{ p{\textwidth}} {\textbf{Definition:} \textit{Individuals’ information is expected to be protected and handled appropriately, wherever the processing takes place.}} \\ 
        \toprule[0.32ex]
        \multirow{2}{50pt}{GDPR} & Article 44 & \textit{General principle for transfers} \\ \cline{2-3}
        & \multicolumn {2}{ p{0.85\textwidth}} {\scriptsize{‘All provisions in this Chapter “Chapter V: Transfers of personal data to third countries or international organisations” shall be applied in order to ensure that the level of protection of natural persons guaranteed by this Regulation is not undermined’.}}  \\
        \toprule
        \multirow{2}{50pt} {PIPEDA} & - & \textit{Personal information transferred across borders} \\ \cline{2-3}
        & \multicolumn {2}{ p{0.85\textwidth}} {\scriptsize{‘Individuals should expect that their information is protected and handled appropriately, regardless of where it's processed’.}}  \\
        \toprule
        \multirow{2}{50pt} {CCPA} & - & \textit{-} \\ \cline{2-3}
        & \multicolumn {2}{ p{0.85\textwidth}} {\scriptsize{N/A 1798.145. (a) ‘The obligations imposed on businesses by this title shall not restrict a business’s ability to: … (6) Collect or sell a consumer’s personal information if every aspect of that commercial conduct takes place wholly outside of California. For purposes of this title, commercial conduct takes place wholly outside of California if the business collected that information while the consumer was outside of California, no part of the sale of the consumer’s personal information occurred in California, and no personal information collected while the consumer was in California is sold. This paragraph shall not permit a business from storing personal information about a consumer when the consumer is in California and then collecting that personal information when the consumer and stored personal information is outside of California’.}}  \\
        \toprule
        \multirow{2}{50pt} {APPS} & App 8 & \textit{Cross-border disclosure of personal information } \\ \cline{2-3}
        & \multicolumn {2}{ p{0.85\textwidth}} {\scriptsize{‘Before an APP entity discloses personal information to an overseas recipient, the entity must take reasonable steps to ensure that the overseas recipient does not breach the APPs in relation to the information’.}}  \\
        \toprule
        \multirow{2}{50pt} {New Zealand} & - & \textit{Does the Privacy Act apply to organisations based overseas?} \\ \cline{2-3}
        & \multicolumn {2}{ p{0.85\textwidth}} {\scriptsize{‘Any organisation operating in New Zealand is required to comply with the Privacy Act, regardless of where the organisation is based’.}}  \\
        \toprule [0.32ex]
    \end{tabular}
    \label{tab:Cross-border Disclosure of Personal Information Principle}
\end{minipage}
\end{table}
\endgroup


\begingroup
\footnotesize
\begin{table}[htbp!]
 \caption{Dealing with Unsolicited Personal Information Principle}
\begin{minipage}{\columnwidth}
        \begin{tabular}{c?l|l}
        \multicolumn{3}{ p{\textwidth}} {\textbf{Definition:} \textit{An organisation must take reasonable steps to manage unsolicited personal information. Unsolicited personal information is ‘personal information received by an entity that has not been requested by that entity.’}} \\ 
        \toprule[0.32ex]
        \multirow{2}{50pt}{GDPR} & - & \textit{-} \\ \cline{2-3}
        & \multicolumn {2}{ p{0.85\textwidth}} {\scriptsize{N/A}}  \\
        \toprule
        \multirow{2}{50pt} {PIPEDA} & - & \textit{-} \\ \cline{2-3}
        & \multicolumn {2}{ p{0.85\textwidth}} {\scriptsize{N/A}}  \\
        \toprule
        \multirow{2}{50pt} {CCPA} & - & \textit{-} \\ \cline{2-3}
        & \multicolumn {2}{ p{0.85\textwidth}} {\scriptsize{N/A}}  \\
        \toprule
        \multirow{2}{50pt} {APPS} & App 4 & \textit{Dealing with unsolicited personal information} \\ \cline{2-3}
        & \multicolumn {2}{ p{0.85\textwidth}} {\scriptsize{4.1 ‘APP 4 outlines the steps an APP entity must take if it receives unsolicited personal information. Unsolicited personal information is personal information received by an entity that has not been requested by that entity’. 
4.2 An APP entity that receives unsolicited personal information must decide whether or not it could have collected the information under APP 3. If not, the entity must destroy or de-identify the information’.
}}  \\
        \toprule
        \multirow{2}{50pt} {New Zealand} & - & \textit{-} \\ \cline{2-3}
        & \multicolumn {2}{ p{0.85\textwidth}} {\scriptsize{N/A}}  \\
        \toprule [0.32ex]
    \end{tabular}
    \label{tab:Dealing with Unsolicited Personal Information Principle}
\end{minipage}

\vspace{5em}


\footnotesize
\caption{Adoption, Use or Disclosure of an identifier Principle}
\begin{minipage}{\columnwidth}
        \begin{tabular}{c?l|l}
        \multicolumn{3}{ p{\textwidth}} {\textbf{Definition:} \textit{Adopting, using, or disclosing a unique identifier that was created for an individual for a different purpose should be prohibited.}} \\ 
        \toprule[0.32ex]
        \multirow{2}{50pt}{GDPR} & Article 87 & \textit{Processing of the national identification number} \\ \cline{2-3}
        & \multicolumn {2}{ p{0.85\textwidth}} {\scriptsize{‘Member States may further determine the specific conditions for the processing of a national identification number or any other identifier of general application. In that case the national identification number or any other identifier of general application shall be used only under appropriate safeguards for the rights and freedoms of the data subject pursuant to this Regulation’.}}  \\
        \toprule
        \multirow{2}{50pt} {PIPEDA} & - & \textit{Guidelines for identification and authentication} \\ \cline{2-3}
        & \multicolumn {2}{ p{0.85\textwidth}} {\scriptsize{‘It is also important to avoid, where ever possible, using numbers such as a driver's licence number or social insurance number as an identifier as they were created for different purposes’.}}  \\
        \toprule
        \multirow{2}{50pt} {CCPA} & - & \textit{-} \\ \cline{2-3}
        & \multicolumn {2}{ p{0.85\textwidth}} {\scriptsize{N/A}}  \\
        \toprule
        \multirow{2}{50pt} {APPS} & App 9 & \textit{Adoption, Use or Disclosure of an identifier} \\ \cline{2-3}
        & \multicolumn {2}{ p{0.85\textwidth}} {\scriptsize{‘An organisation must not adopt, use or disclose a government related identifier unless an exception applies’.}}  \\
        \toprule
        \multirow{2}{50pt} {New Zealand} & Principle 12 & \textit{Unique Identifiers} \\ \cline{2-3}
        & \multicolumn {2}{ p{0.85\textwidth}} {\scriptsize{(2) ‘An agency shall not assign to an individual a unique identifier that, to that agency's knowledge, has been assigned to that individual by another agency, unless those 2 agencies are associated persons within the meaning of section OD7 of the Income Tax Act 2004’.}}  \\
        \toprule [0.32ex]
    \end{tabular}
    \label{tab:doption, Use or Disclosure of an identifier Principle}
\end{minipage}
\end{table}
\endgroup




\begingroup
\footnotesize
\begin{table}[htbp!]
\caption{Right of Individuals to Exercise their Rights}
\begin{minipage}{\columnwidth}
        \begin{tabular}{c?l|l}
        \multicolumn{3}{ p{\textwidth}} {\textbf{Definition:} \textit{Individuals shall have the right to exercise their data protection rights.}} \\ 
        \toprule[0.32ex]
        \multirow{2}{50pt}{GDPR} & Article 12 & \textit{Transparent information, communication and modalities ...} \\ \cline{2-3}
        & \multicolumn {2}{ p{0.85\textwidth}} {\scriptsize{‘The controller shall facilitate the exercise of data subject rights under Articles 15 to 22. In the cases referred to in Article 11(2), the controller shall not refuse to act on the request of the data subject for exercising his or her rights’.}}  \\
        \toprule
        \multirow{2}{50pt} {PIPEDA} & Principle 9 & \textit{Individual Access} \\ \cline{2-3}
        & \multicolumn {2}{ p{0.85\textwidth}} {\scriptsize{‘Help people prepare their request for access to personal information.’
‘Respond to the request as quickly as possible, and no later than 30 days after receiving it’.
}}  \\
        \toprule
        \multirow{2}{50pt} {CCPA} & 1798.130. (a) & \textit{-} \\ \cline{2-3}
        & \multicolumn {2}{ p{0.85\textwidth}} {\scriptsize{‘A business shall: (1) Make available to consumers two or more designated methods for submitting requests for information required to be disclosed pursuant to Sections 1798.110 and 1798.115, including, at a minimum, a toll-free telephone number, and if the business maintains an Internet Web site, a Web site address.    (2) Disclose and deliver the required information to a consumer free of charge within 45 days of receiving a verifiable request from the consumer.’}}  \\
        \toprule
        \multirow{2}{50pt} {APPS} & Your Privacy Rights – App 12 & \textit{Access your personal information - Access to personal information} \\ \cline{2-3}
        & \multicolumn {2}{ p{0.85\textwidth}} {\scriptsize{‘An organisation or agency must be satisfied the request came from you or a person you authorised’.
‘An agency must respond to a request for access to/ [correct] personal information within 30 days’.
‘an entity cannot require an individual to follow a particular procedure, use a designated form or explain the reason for making the request. Any recommended procedure should be regularly reviewed to ensure that it is flexible and facilitates rather than hinders access’.
}}  \\
        \toprule
        \multirow{2}{50pt} {New Zealand} & Principle 6 & \textit{Access to personal information} \\ \cline{2-3}
        & \multicolumn {2}{ p{0.85\textwidth}} {\scriptsize{‘When an agency receives a request for access or correction, it has to do various things. It has to:  decide whether it agrees to the request;   if so, decide how it will provide access or correction;    The agency has to make these decisions and let the requester know as soon as it can. The longest the agency can usually take is 20 working days from the date it receives the request.’}}  \\
        \toprule [0.32ex]
    \end{tabular}

    \label{tab:Right of Individuals to Exercise their Rights}
\end{minipage}

\vspace{5em}


\footnotesize
\caption{Right to be Informed}
\begin{minipage}{\columnwidth}
        \begin{tabular}{c?l|l}
        \multicolumn{3}{ p{\textwidth}} {\textbf{Definition:} \textit{Individuals must be informed about the collection, use and/or disclosure of their personal information.}} \\ 
        \toprule[0.32ex] 
        \multirow{2}{50pt}{GDPR} & Article 13 - 14 & \textit{Information to be provided where personal data are collected/ not obtained ...} \\ \cline{2-3}
        & \multicolumn {2}{ p{0.85\textwidth}} {\scriptsize{‘Where personal data relating to a data subject are collected from the data subject, the controller shall, at the time when personal data are obtained, provide the data subject with all of the [specified] information’, such as information about the controller, the purpose of processing, the recipients of the personal data, and the rights of the data subject…etc.

‘Where personal data have not been obtained from the data subject, the controller shall provide the data subject with the [specified] information.’}}  \\
        \toprule
        \multirow{2}{50pt} {PIPEDA} & Principle 3 & \textit{Consent} \\ \cline{2-3}
        & \multicolumn {2}{ p{0.85\textwidth}} {\scriptsize{‘The knowledge of the individual is required for the collection, use, or disclosure of personal information’.}}  \\
        \toprule
        \multirow{2}{50pt} {CCPA} & 1798.100 (b) & \textit{-} \\ \cline{2-3}
        & \multicolumn {2}{ p{0.85\textwidth}} {\scriptsize{‘A business that collects a consumer’s personal information shall, at or before the point of collection, inform consumers as to the categories of personal information to be collected and the purposes for which the categories of personal information shall be used’.}}  \\
        \toprule
        \multirow{2}{50pt} {APPS} & App 5 & \textit{Notification of the collection of personal information } \\ \cline{2-3}
        & \multicolumn {2}{ p{0.85\textwidth}} {\scriptsize{‘An APP entity that collects personal information about an individual must take reasonable steps either to notify the individual of certain matters or to ensure the individual is aware of those matters’.}}  \\
        \toprule
        \multirow{2}{50pt} {New Zealand} & Principle 3 & \textit{Collection of information from subject} \\ \cline{2-3}
        & \multicolumn {2}{ p{0.85\textwidth}} {\scriptsize{‘When an agency collects personal information from the person the information is about, it has to take reasonable steps to make sure that person knows things like: 
- why it’s being collected
- who will get the information 
- whether the person has to give the information or whether this is voluntary … etc.’.
}}  \\
        \toprule [0.32ex]
    \end{tabular}
   \label{tab:Right to be Informed}
\end{minipage}
\end{table}
\endgroup


\begingroup
\footnotesize
\begin{table}[htbp!]
  \caption{Right of Individuals Access}
\begin{minipage}{\columnwidth}
        \begin{tabular}{c?l|l}
        \multicolumn{3}{ p{\textwidth}} {\textbf{Definition:} \textit{Individuals have the right to access the personal information which an organisation is holding about them.}} \\ 
        \toprule[0.32ex]
        \multirow{2}{50pt}{GDPR} & Article 15 & \textit{Right of access by the data subject} \\ \cline{2-3}
        & \multicolumn {2}{ p{0.85\textwidth}} {\scriptsize{‘The data subject shall have the right to access to the personal data’, and provide information about the processing, such as ‘purposes of processing, categories of personal data concerned, recipients, etc.)’. }}  \\
        \toprule
        \multirow{2}{50pt} {PIPEDA} & Principle 9 & \textit{Individual Access} \\ \cline{2-3}
        & \multicolumn {2}{ p{0.85\textwidth}} {\scriptsize{‘Individuals have a right to access the personal information that an organization holds about them’. ‘When asked, advise people about the personal information about them your organization holds.  Explain how that information is or has been used and to whom it has been disclosed.
 Give people access to their information at minimal or no cost or explain your reasons for not providing access. Providing access can take different forms. For example, you may provide a written or electronic copy of the information or allow the individual to view the information or listen to a recording of the information’.
}}  \\
        \toprule
        \multirow{2}{50pt} {CCPA} & 1798.100. (d) & \textit{-} \\ \cline{2-3}
        & \multicolumn {2}{ p{0.85\textwidth}} {\scriptsize{‘A business that receives a verifiable consumer request from a consumer to access personal information shall promptly take steps to disclose and deliver, free of charge to the consumer …  the categories and specific pieces of personal information the business has collected’.}}  \\
        \toprule
        \multirow{2}{50pt} {APPS} & App 12 & \textit{Access to personal information} \\ \cline{2-3}
        & \multicolumn {2}{ p{0.85\textwidth}} {\scriptsize{‘APP 12 requires an APP entity that holds personal information about an individual to give the individual access to that information on request’ ‘The manner of access may, for example, be by email, by phone, in person, hard copy, or an electronic record’.
 ‘An agency cannot impose upon an individual any charge for providing access to personal information’.}}  \\
        \toprule
        \multirow{2}{50pt} {New Zealand} & Principle 6 & \textit{Access to personal information} \\ \cline{2-3}
        & \multicolumn {2}{ p{0.85\textwidth}} {\scriptsize{‘People usually have a right to ask for access to personal information that identifies them’ ‘… a public sector agency shall not require the payment, by or on behalf of any individual who wishes to make an information privacy request’.}}  \\
        \toprule [0.32ex]
    \end{tabular}
  
   \label{tab:Right of Individuals Access}
\end{minipage}

\vspace{5em}


\footnotesize
\caption{Right to Rectification}
\begin{minipage}{\columnwidth}
        \begin{tabular}{c?l|l}
        \multicolumn{3}{ p{\textwidth}} {\textbf{Definition:} \textit{Individuals shall have the right to ask for correction of their personal data.}} \\ 
        \toprule[0.32ex]
        \multirow{2}{50pt}{GDPR} & Article 16 & \textit{Right to rectification } \\ \cline{2-3}
        & \multicolumn {2}{ p{0.85\textwidth}} {\scriptsize{‘The data subject shall have the right to obtain from the controller without undue delay the rectification of inaccurate personal data concerning him or her’.}}  \\
        \toprule
        \multirow{2}{50pt} {PIPEDA} & Principle 9 & \textit{Individual Access } \\ \cline{2-3}
        & \multicolumn {2}{ p{0.85\textwidth}} {\scriptsize{‘Individuals have the right to challenge the accuracy and completeness of the information, and have that information amended as appropriate’.}}  \\
        \toprule
        \multirow{2}{50pt} {CCPA} & - & \textit{-} \\ \cline{2-3}
        & \multicolumn {2}{ p{0.85\textwidth}} {\scriptsize{N/A}}  \\
        \toprule
        \multirow{2}{50pt} {APPS} & App 13 & \textit{Correction of personal information } \\ \cline{2-3}
        & \multicolumn {2}{ p{0.85\textwidth}} {\scriptsize{‘Australian privacy law gives you the right to correct the personal information an organization or agency holds about you if it is: inaccurate, out of date, incomplete, irrelevant, misleading’.}}  \\
        \toprule
        \multirow{2}{50pt} {New Zealand} & Principle 7 & \textit{Correction of personal information } \\ \cline{2-3}
        & \multicolumn {2}{ p{0.85\textwidth}} {\scriptsize{‘People have a right to ask the agency to correct information about themselves, if they think it is wrong’.}}  \\
        \toprule [0.32ex]
    \end{tabular}
    
   \label{tab:Right to Rectification}
\end{minipage}
\end{table}
\endgroup


\begingroup
\footnotesize
\begin{table}[htbp!]
 \caption{Right to Erasure}
\begin{minipage}{\columnwidth}
        \begin{tabular}{c?l|l}
        \multicolumn{3}{ p{\textwidth}} {\textbf{Definition:} \textit{Individuals shall have the right to ask for deletion of their personal data.}} \\ 
        \toprule[0.32ex]
        \multirow{2}{50pt}{GDPR} & Article 17 & \textit{Right to erasure (‘right to be forgotten’)} \\ \cline{2-3}
        & \multicolumn {2}{ p{0.85\textwidth}} {\scriptsize{ ‘The data subject shall have the right to obtain from the controller the erasure of personal data concerning him or her without undue delay and the controller shall have the obligation to erase personal data without undue delay where [one of the specified grounds applies]’.}}  \\
        \toprule
        \multirow{2}{50pt} {PIPEDA} & Principle 5 - Principle 6 & \textit{Principle 5 - Limiting Use, Disclosure, and Retention
Principle 6- Accuracy} \\ \cline{2-3}
        & \multicolumn {2}{ p{0.85\textwidth}} {\scriptsize{ ‘Destroy, erase or anonymise any personal information that your organization no longer needs’.
‘When an individual successfully demonstrates the inaccuracy or incompleteness of personal information, the organization shall amend the information as required. Depending upon the nature of the information challenged, amendment involves the correction, deletion, or addition of information’\cite{InterpretationBulletinAccuracy} .
}}  \\
        \toprule
        \multirow{2}{50pt} {CCPA} & 1798.105. (a) & \textit{-} \\ \cline{2-3}
        & \multicolumn {2}{ p{0.85\textwidth}} {\scriptsize{‘A consumer shall have the right to request that a business delete any personal information about the consumer which the business has collected from the consumer’. }}  \\
        \toprule
        \multirow{2}{50pt} {APPS} & - & \textit{Australian entities and the EU General Data Protection Regulation (GDPR)} \\ \cline{2-3}
        & \multicolumn {2}{ p{0.85\textwidth}} {\scriptsize{‘There is no equivalent ‘right to erasure’ under the Privacy Act, however APP 11.2 requires an APP entity that holds personal information to take reasonable steps to destroy the information or to ensure it is de-identified if the information is no longer needed for any purpose permitted under the Privacy Act’ \cite{AppsEU}.  }}  \\
        \toprule
        \multirow{2}{50pt} {New Zealand} & Principle 9 & \textit{Agency not to keep personal information for longer than necessary } \\ \cline{2-3}
        & \multicolumn {2}{ p{0.85\textwidth}} {\scriptsize{ There is no equivalent ‘right to erasure’ under the New Zealand’s Privacy Act, however, ‘an agency that holds personal information must not keep that information for longer than is necessary for the purposes for which the information may be lawfully used’.}}  \\
        \toprule [0.32ex]
    \end{tabular}
    \label{tab:Right to Erasure}
\end{minipage}

\vspace{5em}


\footnotesize
\caption{Right to Restriction of Processing}
\begin{minipage}{\columnwidth}
        \begin{tabular}{c?l|l}
        \multicolumn{3}{ p{\textwidth}} {\textbf{Definition:} \textit{Individuals shall have the right to restrict the processing of their personal data.}} \\ 
        \toprule[0.32ex]
        \multirow{2}{50pt}{GDPR} & Article 18 & \textit{Right to restriction of processing} \\ \cline{2-3}
        & \multicolumn {2}{ p{0.85\textwidth}} {\scriptsize{‘The data subject shall have the right to obtain from the controller restriction of processing where [one of the specified cases applies]’.}}  \\
        \toprule
        \multirow{2}{50pt} {PIPEDA} & - & \textit{-} \\ \cline{2-3}
        & \multicolumn {2}{ p{0.85\textwidth}} {\scriptsize{N/A}}  \\
        \toprule
        \multirow{2}{50pt} {CCPA} & - & \textit{-} \\ \cline{2-3}
        & \multicolumn {2}{ p{0.85\textwidth}} {\scriptsize{N/A}}  \\
        \toprule
        \multirow{2}{50pt} {APPS} & - & \textit{-} \\ \cline{2-3}
        & \multicolumn {2}{ p{0.85\textwidth}} {\scriptsize{N/A}}  \\
        \toprule
        \multirow{2}{50pt} {New Zealand} & - & \textit{-} \\ \cline{2-3}
        & \multicolumn {2}{ p{0.85\textwidth}} {\scriptsize{N/A}}  \\
        \toprule [0.32ex]
    \end{tabular}
    
   \label{tab:Right to Restriction of Processing}
\end{minipage}
\end{table}
\endgroup


\begingroup
\footnotesize
\begin{table}[htbp!]
\caption{Right to Object}
\begin{minipage}{\columnwidth}
        \begin{tabular}{c?l|l}
        \multicolumn{3}{ p{\textwidth}} {\textbf{Definition:} \textit{Individuals shall have the right to object to the processing of their personal data.}} \\ 
        \toprule[0.32ex]
        \multirow{2}{50pt}{GDPR} & Article 21 & \textit{Right to Object} \\ \cline{2-3}
        & \multicolumn {2}{ p{0.85\textwidth}} {\scriptsize{‘The data subject shall have the right to object, on grounds relating to his or her particular situation, at any time to processing of personal data concerning him or her’.}}  \\
        \toprule
        \multirow{2}{50pt} {PIPEDA} & Principle 3 & \textit{Consent} \\ \cline{2-3}
        & \multicolumn {2}{ p{0.85\textwidth}} {\scriptsize{There is no specific right to object to processing under the PIPEDA, but it can be inferred from principle 3 that ‘individuals can withdraw consent at any time, subject to legal or contractual restrictions and reasonable notice, and organizations must inform individual of the implications of withdrawal’. }}  \\
        \toprule
        \multirow{2}{50pt} {CCPA} & 1798.120. (a) & \textit{-} \\ \cline{2-3}
        & \multicolumn {2}{ p{0.85\textwidth}} {\scriptsize{‘A consumer shall have the right, at any time, to direct a business that sells personal information about the consumer to third parties not to sell the consumer’s personal information. This right may be referred to as the right to opt out’.}}  \\
        \toprule
        \multirow{2}{50pt} {APPS} & Chapter B & \textit{B.51 Key concepts} \\ \cline{2-3}
        & \multicolumn {2}{ p{0.85\textwidth}} {\scriptsize{There is no specific right to object to processing under the APPs, but ‘An individual may withdraw their consent at any time’ [3]}}  \\
        \toprule
        \multirow{2}{50pt} {New Zealand} & - & \textit{When does a privacy waiver expire?} \\ \cline{2-3}
        & \multicolumn {2}{ p{0.85\textwidth}} {\scriptsize{N/A
The New Zealand’s Privacy Act does not contain an equivalent right to ‘right to object’. However, ‘an individual might at any point withdraw that consent and the agency has to respond accordingly’.
}}  \\
        \toprule [0.32ex]
    \end{tabular}
  \label{tab:Right to Object}
\end{minipage}

\vspace{5em}


\footnotesize
\caption{Right to Object to Marketing}
\begin{minipage}{\columnwidth}
        \begin{tabular}{c?l|l}
        \multicolumn{3}{ p{\textwidth}} {\textbf{Definition:} \textit{Individuals shall have the right to reject the processing of their personal data for direct marketing purposes.}} \\ 
        \toprule[0.32ex]
        \multirow{2}{50pt}{GDPR} & Article 21 & \textit{Right to object } \\ \cline{2-3}
        & \multicolumn {2}{ p{0.85\textwidth}} {\scriptsize{2.   ‘Where personal data are processed for direct marketing purposes, the data subject shall have the right to object at any time to processing of personal data concerning him or her for such marketing, which includes profiling to the extent that it is related to such direct marketing’.
3.   ‘Where the data subject objects to processing for direct marketing purposes, the personal data shall no longer be processed for such purposes’.
}}  \\
        \toprule
        \multirow{2}{50pt} {PIPEDA} & - & \textit{-} \\ \cline{2-3}
        & \multicolumn {2}{ p{0.85\textwidth}} {\scriptsize{N/A}}  \\
        \toprule
        \multirow{2}{50pt} {CCPA} & 1798.120. (c) & \textit{-} \\ \cline{2-3}
        & \multicolumn {2}{ p{0.85\textwidth}} {\scriptsize{It is not mentioned explicitly but could be inferred from 1798.120. (c), ‘a business that has received direction from a consumer not to sell the consumer’s personal information or, in the case of a minor consumer’s personal information has not received consent to sell the minor consumer’s personal information shall be prohibited, pursuant to paragraph (4) of subdivision (a) of Section 1798.135, from selling the consumer’s personal information after its receipt of the consumer’s direction, unless the consumer subsequently provides express authorization for the sale of the consumer’s personal information’.}}  \\
        \toprule
        \multirow{2}{50pt} {APPS} & App 7 & \textit{Direct marketing} \\ \cline{2-3}
        & \multicolumn {2}{ p{0.85\textwidth}} {\scriptsize{“An organisation must not use or disclose the personal information that it holds about an individual for the purpose of direct marketing (APP 7.1)
7.2 and 7.3 require an organisation to provide a simple means by which an individual can request not to receive direct marketing communications (also known as ‘opting out’).”
}}  \\
        \toprule
        \multirow{2}{50pt} {New Zealand} & - & \textit{-} \\ \cline{2-3}
        & \multicolumn {2}{ p{0.85\textwidth}} {\scriptsize{N/A}}  \\
        \toprule [0.32ex]
    \end{tabular}
   \label{tab:Right to Object to Marketing}
\end{minipage}
\end{table}
\endgroup


\begingroup
\footnotesize
\begin{table}[htbp!]
\caption{Right to Data Portability}
\begin{minipage}{\columnwidth}
        \begin{tabular}{c?l|l}
        \multicolumn{3}{ p{\textwidth}} {\textbf{Definition:} \textit{Individuals shall have the right to receive their personal data in a readily useable format that allows the Individual to transmit the information to another organisation.}} \\ 
        \toprule[0.32ex]
        \multirow{2}{50pt}{GDPR} & Article 20 & \textit{Right to data Portability } \\ \cline{2-3}
        & \multicolumn {2}{ p{0.85\textwidth}} {\scriptsize{‘The data subject shall have the right to receive the personal data concerning him or her, which he or she has provided to a controller, in a structured, commonly used and machine-readable format and have the right to transmit those data to another controller without hindrance from the controller to which the personal data have been provided’.}}  \\
        \toprule
        \multirow{2}{50pt} {PIPEDA} & - & \textit{-} \\ \cline{2-3}
        & \multicolumn {2}{ p{0.85\textwidth}} {\scriptsize{N/A}}  \\
        \toprule
        \multirow{2}{50pt} {CCPA} & 1798.100. (d) & \textit{-} \\ \cline{2-3}
        & \multicolumn {2}{ p{0.85\textwidth}} {\scriptsize{‘A business that receives a verifiable consumer request from a consumer to access personal information shall promptly take steps to disclose and deliver, free of charge to the consumer, the personal information required by this section. The information may be delivered by mail or electronically, and if provided electronically, the information shall be in a portable and, to the extent technically feasible, in a readily useable format that allows the consumer to transmit this information to another entity without hindrance’.}}  \\
        \toprule
        \multirow{2}{50pt} {APPS} & - & \textit{Australian entities and the EU General Data Protection Regulation (GDPR)} \\ \cline{2-3}
        & \multicolumn {2}{ p{0.85\textwidth}} {\scriptsize{N/A
‘The Privacy Act does not include an equivalent right to ‘data portability’ or ‘right to object’. However, individuals do have a right to request access to, and correction of, their personal information under APPs 12 and 13. In giving access under APP 12, where reasonable and practicable, an entity must give access in the manner requested by the individual. For more information, see Chapters 12 and 13 of the APP guidelines’ \cite{AppsEU}. }}  \\
        \toprule
        \multirow{2}{50pt} {New Zealand} & - & \textit{-} \\ \cline{2-3}
        & \multicolumn {2}{ p{0.85\textwidth}} {\scriptsize{N/A}}  \\
        \toprule [0.32ex]
    \end{tabular}
    \label{tab:Right to Data Portability}
\end{minipage}
\end{table}


\footnotesize
\begin{table}[htbp!]
\caption{Right to Object to Automated Decision-Making}
\begin{minipage}{\columnwidth}
        \begin{tabular}{c?l|l}
        \multicolumn{3}{ p{\textwidth}} {\textbf{Definition:} \textit{Individuals must have the right to object to a decision that is based solely on automated processing.}} \\ 
        \toprule[0.32ex]
        \multirow{2}{50pt}{GDPR} & Article 22 & \textit{Automated individual decision-making, including profiling } \\ \cline{2-3}
        & \multicolumn {2}{ p{0.85\textwidth}} {\scriptsize{‘The data subject shall have the right not to be subject to a decision based solely on automated processing, including profiling, which produces legal effects concerning him or her or similarly significantly affects him or her’.}}  \\
        \toprule
        \multirow{2}{50pt} {PIPEDA} & - & \textit{-} \\ \cline{2-3}
        & \multicolumn {2}{ p{0.85\textwidth}} {\scriptsize{N/A}}  \\
        \toprule
        \multirow{2}{50pt} {CCPA} & - & \textit{-} \\ \cline{2-3}
        & \multicolumn {2}{ p{0.85\textwidth}} {\scriptsize{N/A}}  \\
        \toprule
        \multirow{2}{50pt} {APPS} & - & \textit{-} \\ \cline{2-3}
        & \multicolumn {2}{ p{0.85\textwidth}} {\scriptsize{N/A}}  \\
        \toprule
        \multirow{2}{50pt} {New Zealand} & - & \textit{-} \\ \cline{2-3}
        & \multicolumn {2}{ p{0.85\textwidth}} {\scriptsize{N/A}}  \\
        \toprule [0.32ex]
    \end{tabular}
    
    \label{tab:Right to Object to Automated Decision-Making}
\end{minipage}
\end{table}
\endgroup


\begingroup
\footnotesize
\begin{table}[htbp!]
\caption{Right to Withdraw Consent}
\begin{minipage}{\columnwidth}
        \begin{tabular}{c?l|l}
        \multicolumn{3}{ p{\textwidth}} {\textbf{Definition:} \textit{Individuals shall have the right to withdraw consent at any time.}} \\ 
        \toprule[0.32ex]
        \multirow{2}{50pt}{GDPR} & Article 7 & \textit{Conditions for consent} \\ \cline{2-3}
        & \multicolumn {2}{ p{0.85\textwidth}} {\scriptsize{‘The data subject shall have the right to withdraw his or her consent at any time. The withdrawal of consent shall not affect the lawfulness of processing based on consent before its withdrawal. Prior to giving consent, the data subject shall be informed thereof. It shall be as easy to withdraw as to give consent’.}}  \\
        \toprule
        \multirow{2}{50pt} {PIPEDA} & Principle 3 & \textit{Consent } \\ \cline{2-3}
        & \multicolumn {2}{ p{0.85\textwidth}} {\scriptsize{‘Individuals can withdraw consent at any time, subject to legal or contractual restrictions and reasonable notice, and you must inform individual of the implications of withdrawal’. }}  \\
        \toprule
        \multirow{2}{50pt} {CCPA} & 1798.120. (a) & \textit{-} \\ \cline{2-3}
        & \multicolumn {2}{ p{0.85\textwidth}} {\scriptsize{‘A consumer shall have the right, at any time, to direct a business that sells personal information about the consumer to third parties not to sell the consumer’s personal information. This right may be referred to as the right to opt out’. }}  \\
        \toprule
        \multirow{2}{50pt} {APPS} & Chapter B & \textit{B.51 Key concepts} \\ \cline{2-3}
        & \multicolumn {2}{ p{0.85\textwidth}} {\scriptsize{‘An individual may withdraw their consent at any time, and this should be an easy and accessible process. Once an individual has withdrawn consent, an APP entity can no longer rely on that past consent for any future use or disclosure of the individual’s personal information’ \cite{KeyconceptsB}}}  \\
        \toprule
        \multirow{2}{50pt} {New Zealand} & - & \textit{When does a privacy waiver expire?} \\ \cline{2-3}
        & \multicolumn {2}{ p{0.85\textwidth}} {\scriptsize{‘An individual might at any point withdraw that consent and the agency has to respond accordingly’ \cite{Whendoesaprivacywaiverexpire}. }}  \\
        \toprule [0.32ex]
    \end{tabular}
   \label{tab:Right to Withdraw Consent}
\end{minipage}

\vspace{5em}


\footnotesize
\caption{Right to Complain}
\begin{minipage}{\columnwidth}
        \begin{tabular}{c?l|l}
        \multicolumn{3}{ p{\textwidth}} {\textbf{Definition:} \textit{Every data subject shall have the right to lodge a complaint.}} \\ 
        \toprule[0.32ex]
        \multirow{2}{50pt}{GDPR} & Article 14 & \textit{Transparent information, communication and modalities for the exercise of the rights ...} \\ \cline{2-3}
        & \multicolumn {2}{ p{0.85\textwidth}} {\scriptsize{‘If the controller does not take action on the request of the data subject, the controller shall inform the data subject without delay and at the latest within one month of receipt of the request of the reasons for not taking action and on the possibility of lodging a complaint with a supervisory authority and seeking a judicial remedy’.}}  \\
        \toprule
        \multirow{2}{50pt} {PIPEDA} & Principle 10 & \textit{Challenging Compliance - File a formal privacy complaint } \\ \cline{2-3}
        & \multicolumn {2}{ p{0.85\textwidth}} {\scriptsize{‘An individual must be able to challenge your organization’s compliance with the fair information principles. They should address their challenge to the person in your organization who is accountable for compliance with PIPEDA’.
‘If you can’t resolve the issue directly with the organization, you may decide to file a formal complaint’.}}  \\
        \toprule
        \multirow{2}{50pt} {CCPA} & 1798.150. (a) & \textit{-} \\ \cline{2-3}
        & \multicolumn {2}{ p{0.85\textwidth}} {\scriptsize{‘(1) Any consumer whose nonencrypted or nonredacted personal information is subject to an unauthorized access and exfiltration, theft, or disclosure as a result of the business’ violation of the duty to implement and maintain reasonable security procedures and practices appropriate to the nature of the information to protect the personal information may institute a civil action for any of [specified conditions]’.
..... ‘(2) A consumer bringing an action as defined in paragraph (1) of subdivision (c) shall notify the Attorney General within 30 days that the action has been filed.
(3) The Attorney General, upon receiving such notice shall, within 30 days, do one of the specified steps’.
}}  \\
        \toprule
        \multirow{2}{50pt} {APPS} & - & \textit{Lodge a privacy complaint with us} \\ \cline{2-3}
        & \multicolumn {2}{ p{0.85\textwidth}} {\scriptsize{‘If you’re concerned your personal information has been mishandled, you first need to complain to the organisation or agency you think has mishandled it. If they don’t respond to your complaint within 30 days or you’re not happy with their response, you can lodge a complaint with us’.}}  \\
        \toprule
        \multirow{2}{50pt} {New Zealand} & Your Rights & \textit{Making a complaint} \\ \cline{2-3}
        & \multicolumn {2}{ p{0.85\textwidth}} {\scriptsize{‘If you’re unable to resolve a privacy dispute, you can make a complaint to us. But, before you make a complaint to us, contact the agency and let it know what the problem is. Agencies can often sort out problems more quickly without the need for us to get involved’.}}  \\
        \toprule [0.32ex]
    \end{tabular}
   \label{tab:Right to Complain}
\end{minipage}
\end{table}
\endgroup

\clearpage


\begingroup
\footnotesize
\begin{table}[ht!]
\caption{Right of individuals not to  be Discriminated}
\begin{minipage}{\columnwidth}
        \begin{tabular}{c?l|l}
        \multicolumn{3}{ p{\textwidth}} {\textbf{Definition:} \textit{A business shall not discriminate against a consumer because the consumer exercised any of their rights.}} \\ 
        \toprule[0.32ex]
        \multirow{2}{50pt}{GDPR} & - & \textit{-} \\ \cline{2-3}
        & \multicolumn {2}{ p{0.85\textwidth}} {\scriptsize{N/A}}  \\
        \toprule
        \multirow{2}{50pt} {PIPEDA} & - & \textit{-} \\ \cline{2-3}
        & \multicolumn {2}{ p{0.85\textwidth}} {\scriptsize{N/A}}  \\
        \toprule
        \multirow{2}{50pt} {CCPA} & 1798.130. (a) & \textit{-} \\ \cline{2-3}
        & \multicolumn {2}{ p{0.85\textwidth}} {\scriptsize{“1798.125. (a) (1) A business shall not discriminate against a consumer because the consumer exercised any of the consumer’s rights under this title.”}}  \\
        \toprule
        \multirow{2}{50pt} {APPS} & - & \textit{-} \\ \cline{2-3}
        & \multicolumn {2}{ p{0.85\textwidth}} {\scriptsize{N/A}}  \\
        \toprule
        \multirow{2}{50pt} {New Zealand} & - & \textit{-} \\ \cline{2-3}
        & \multicolumn {2}{ p{0.85\textwidth}} {\scriptsize{N/A}}  \\
        \toprule [0.32ex]
    \end{tabular}
  \label{tab:Right of not being Discriminated}
\end{minipage}
\end{table}
\endgroup



\section{Privacy by Design Schemes}\label{Appendix:Privacy by Design Schemes}

This section presents various Privacy by Design (PbD) schemes (e.g., privacy principles, strategies, guidelines, and patterns) that developed previously by different researchers.
\subsection{Privacy Principles}\label{Appendix:Privacy Principles}


\subsubsection{\textbf{Privacy by Design Principles by Cate (2010)} Fair Information Practice Principles (FIPPs) \cite{Cate2006}:} \label{Appendix:Privacy by Design Principles by Cate (2010)}

\textbf{01. Notice / Awareness} Consumers should be given notice of an entity’s information practices before any personal information is collected from them. This requires that companies explicitly notify some or all of the following:
\begin{itemize}
    \item Identification of the entity collecting the data;
    \item Identification of the uses to which the data will be put;
    \item Identification of any potential recipients of the data; 
    \item The nature of the data collected and the means by which it is collected;
    \item Whether the provision of the requested data is voluntary or required; 
    \item The steps taken by the data collector to ensure the confidentiality, integrity and quality of the data.

\end{itemize}

\textbf{02. Choice / Consent Choice} and consent in an on-line information-gathering sense means giving consumers options to control how their data is used. Specifically, choice relates to secondary uses of information beyond the immediate needs of the information collector to complete the consumer’s transaction. The two typical types of choice models are ’opt- in’ or ’opt-out.’ The ’opt-in’ method requires that consumers affirmatively give permission for their information to be used for other purposes. Without the consumer taking these affirmative steps in an ’opt-in’ system, the information gatherer assumes that it cannot use the information for any other purpose. The ’opt-out’ method requires consumers to affirmatively decline permission for other uses. Without the consumer taking these affirmative steps in an ’opt-out’ system, the information gatherer assumes that it can use the consumer’s information for other purposes. Each of these systems can be designed to allow an individual consumer to tailor the information gatherer’s use of the information to fit their preferences by checking boxes to grant or deny permission for specific purposes rather than using a simple "all or nothing" method.

\textbf{03. Access / Participation} Access as defined in the Fair Information Practice Principles includes not only a consumer’s ability to view the data collected, but also to verify and contest its accuracy. This access must be inexpensive and timely in order to be useful to the consumer.

\textbf{04. Integrity / Security} Information collectors should ensure that the data they collect is accurate and secure. They can improve the integrity of data by cross-referencing it with only reputable databases and by providing access for the consumer to verify it. Information collectors can keep their data secure by protecting against both internal and external security threats. They can limit access within their company to only necessary employees to protect against internal threats, and they can use encryption and other computer-based security systems to stop outside threats

\textbf{05. Enforcement / Redress} In order to ensure that companies follow the Fair Information Practice Principles, there must be enforcement measures. The FTC identified three types of enforcement measures: self-regulation by the information collectors or an appointed regulatory body; private remedies that give civil causes of action for individuals whose information has been misused to sue violators; and government enforcement that can include civil and criminal penalties levied by the government.

\subsubsection{\textbf{Privacy by Design Principles by Ann Cavoukian}} Cavoukian (2010) has proposed seven Privacy by Design foundation principles \cite{Cavoukian2010}: \label{Appendix:Privacy by Design Principles by Ann Cavoukian (2010)}

\textbf{01. Proactive not Reactive; Preventative not Remedial} The Privacy by Design (PbD) approach is characterized by proactive rather than reactive measures. It anticipates and prevents privacy invasive events before they happen. PbD does not wait for privacy risks to materialize, nor does it offer remedies for resolving privacy infractions once they have occurred - it aims to prevent them from occurring. In short, Privacy by Design comes before-the-fact, not after.

\textbf{02.Privacy as the Default Setting} We can all be certain of one thing - the default rules! Privacy by Design seeks to deliver the maximum degree of privacy by ensuring that personal data are automatically protected in any given IT system or business practice. If an individual does nothing, their privacy still remains intact. No action is required on the part of the individual to protect their privacy - it is built into the system, by default.
03. Privacy Embedded into Design Privacy by Design is embedded into the design and architecture of IT systems and business practices. It is not bolted on as an add-on, after the fact. The result is that privacy becomes an essential component of the core functionality being delivered. Privacy is integral to the system, without diminishing functionality.

\textbf{03.Privacy Embedded into Design} Privacy by Design is embedded into the design and architecture of IT systems and business practices. It is not bolted on as an add-on, after the fact. The result is that privacy becomes an essential component of the core functionality being delivered. Privacy is integral to the system, without diminishing functionality.

\textbf{04. Full Functionality-Positive - Sum, not Zero-Sum} Privacy by Design seeks to accommodate all legitimate interests and objectives in a positive-sum “win-win” manner, not through a dated, zero-sum approach, where unnecessary trade-offs are made. Privacy by Design avoids the pretense of false dichotomies, such as privacy vs. security, demonstrating that it is possible to have both.

\textbf{05. End-to-End Security - Full Lifecycle Protection} Privacy by Design, having been em- bedded into the system prior to the first element of information being collected, extends securely throughout the entire lifecycle of the data involved - strong security measures are essential to privacy, from start to finish. This ensures that all data are securely retained, and then securely destroyed at the end of the process, in a timely fashion. Thus, Privacy by Design ensures cradle to grave, secure lifecycle management of information, end-to-end.

\textbf{06. Visibility and Transparency - Keep it Open} Privacy by Design seeks to assure all stake- holders that whatever the business practice or technology involved, it is in fact, operating according to the stated promises and objectives, subject to independent verification. Its component parts and operations remain visible and transparent, to users and providers alike. Remember, trust but verify.

\textbf{07. Respect for User Privacy - Keep it User-Centric} Above all, Privacy by Design requires architects and operators to keep the interests of the individual uppermost by offering such measures as strong privacy defaults, appropriate notice, and empowering user-friendly options. Keep it user-centric.

\subsubsection{\textbf{ISO 29100 Privacy framework }} ISO 29100 has proposed 11 privacy principles \cite{Iso/Iec270322012}: \label{Appendix:ISO 29100 Privacy framework} 

\textbf{01. Consent and choice} Adhering to the consent principle means: 

\begin{itemize}
    \item  presenting to the PII principal the choice whether or not to allow the processing of their PII except where the PII principal cannot freely withhold consent or where applicable law specifically allows the processing of PII without the natural personâĂŹs consent. The PII principal’s choice must be given freely, specific and on a knowledgeable basis;
    
    \item obtaining the opt-in consent of the PII principal for collecting or otherwise processing sensitive PII except where applicable law allows the processing of sensitive PII without the natural personâĂŹs consent;
    
    \item informing PII principals, before obtaining consent, about their rights under the individual participation and access principle;
    
    \item providing PII principals, before obtaining consent, with the information indicated by the openness, transparency and notice principle; and
    
   \item explaining to PII principals the implications of granting or withholding consent.

\end{itemize}

\textbf{02. Purpose legitimacy and specification} Adhering to the purpose legitimacy and specification principle means:

\begin{itemize}
     
\item ensuring that the purpose(s) complies with applicable law and relies on a permissible legal basis;
\item communicating the purpose(s) to the PII principal before the time the information is collected or used for the first time for a new purpose;
\item using language for this specification which is both clear and appropriately adapted to the circumstances; and
\item if applicable, giving sufficient explanations for the need to process sensitive PII.

\end{itemize}

\textbf{03. Collection limitation} Adhering to the collection limitation principle means:

\begin{itemize}
\item limiting the collection of PII to that which is within the bounds of applicable law and strictly necessary for the specified purpose(s).
\end{itemize}

\textbf{04. Data minimisation} Adhering to the Data Minimisation principle means designing and implementing data processing procedures and ICT systems in such a way as to: 

\begin{itemize}
\item minimise the PII which is processed and the number of privacy stakeholders and people to whom PII is disclosed or who have access to it;
\item  ensure adoption of a âĂĲneed-to-knowâĂİ principle, i.e. one should be given access only to the PII which is necessary for the conduct of his/her official duties in the framework of the legitimate purpose of the PII processing;
\item  use or offer as default options, wherever possible, interactions and transactions which do not involve the identification of PII principals, reduce the observability of their behaviour and limit the linkability of the PII collected; and
\item  delete and dispose of PII whenever the purpose for PII processing has expired, there are no legal requirements to keep the PII or whenever it is practical to do so.
\end{itemize}

\textbf{05. Use, retention and disclosure limitation} Adhering to the use, retention and disclosure limitation principle means:

\begin{itemize}
\item Limiting the use, retention and disclosure (including transfer) of PII to that which is necessary in order to fulfil specific, explicit and legitimate purposes;
\item  Limiting the use of PII to the purposes specified by the PII controller prior to collection, unless a different purpose is explicitly required by applicable law;
\item  Retaining PII only as long as necessary to fulfil the stated purposes, and thereafter securely destroying or anonymising it; and
\item  Locking (i.e. archiving, securing and exempting the PII from further processing) any PII when and for as long as the stated purposes have expired, but where retention is required by applicable laws.
\end{itemize}

\textbf{06. Accuracy and quality} Adhering to the accuracy and quality principle means: 

\begin{itemize}
\item  ensuring that the PII processed is accurate, complete, up-to-date (unless there is a legitimate basis for keeping outdated data), adequate and relevant for the purpose of use;
\item  ensuring the reliability of PII collected from a source other than from the PII principal before it is processed;
\item  verifying, through appropriate means, the validity and correctness of the claims made by the PII principal prior to making any changes to the PII (in order to ensure that the changes are properly authorized), where it is appropriate to do so;
\item  establishing PII collection procedures to help ensure accuracy and quality; and
\item  establishing control mechanisms to periodically check the accuracy and quality of collected and stored PII.

\end{itemize}

\textbf{07. Openness, transparency and notice} Adhering to the openness, transparency and notice principle means:

\begin{itemize}

\item providing PII principals with clear and easily accessible information about the PII controller âĂŹs policies, procedures and practices with respect to the processing of PII;
\item including in notices the fact that PII is being processed, the purpose for which this is done, the types of privacy stakeholders to whom the PII might be disclosed, and the identity of the PII controller including information on how to contact the PII controller;
\item disclosing the choices and means offered by the PII controller to PII principals for the purposes of limiting the processing of, and for accessing, correcting and removing their information; and
\item giving notice to the PII principals when major changes in the PII handling procedures occur.
\end{itemize}

\textbf{08. Individual participation and access}; Adhering to the individual participation and access principle means:

\begin{itemize}

\item giving PII principals the ability to access and review their PII, provided their identity is first authenticated with an appropriate level of assurance and such access is not prohibited by applicable law;
\item allowing PII principals to challenge the accuracy and completeness of the PII and have it amended, corrected or removed as appropriate and possible in the specific context;
\item providing any amendment, correction or removal to PII processors and third parties to whom personal data had been disclosed, where they are known; and
\item establishing procedures to enable PII principals to exercise these rights in a simple, fast and efficient way, which does not entail undue delay or cost.
\end{itemize}

\textbf{09. Accountability} The processing of PII entails a duty of care and the adoption of concrete and practical measures for its protection. Adhering to the accountability principle means: 

 \begin{itemize}
\item documenting and communicating as appropriate all privacy-related policies, procedures and practices;

\item assigning to a specified individual within the organization (who might in turn delegate to others in the organization as appropriate) the task of implementing the privacy-related policies, procedures and practices;
\item when transferring PII to third parties, ensuring that the third party recipient will be bound to provide an equivalent level of privacy protection through contractual or other means such as mandatory internal policies (applicable law can contain additional requirements regarding international data transfers);
\item providing suitable training for the personnel of the PII controller who will have access to PII;
\item setting up efficient internal complaint handling and redress procedures for use by PII principals;
\item informing PII principals about privacy breaches that can lead to substantial damage to them (unless prohibited, e.g., while working with law enforcement) as well as the measures taken for resolution;
\item notifying all relevant privacy stakeholders about privacy breaches as required in some jurisdictions (e.g., the data protection authorities) and depending on the level of risk;
\item allowing an aggrieved PII principal access to appropriate and effective sanctions and/or remedies, such as rectification, expungement or restitution if a privacy breach has occurred; and
\item considering procedures for compensation for situations in which it will be difficult or impossible to bring the natural person âĂŹs privacy status back to a position as if nothing had occurred.

\end{itemize}

\textbf{10. Information security} Adhering to the information security principle means: 

\begin{itemize}
\item protecting PII under its authority with appropriate controls at the operational, functional and strategic level to ensure the integrity, confidentiality and availability of the PII, and protect it against risks such as unauthorized access, destruction, use, modification, disclosure or loss throughout the whole of its life cycle;
\item choosing PII processors that provide sufficient guarantees with regard to organizational, physical and technical controls for the processing of PII and ensuring compliance with these controls;
\item basing these controls on applicable legal requirements, security standards, the results of systematic security risk assessments as described in ISO 31000, and the results of a cost/benefit analysis;
\item implementing controls in proportion to the likelihood and severity of the potential consequences, the sensitivity of the PII, the number of PII principals that might be affected, and the context in which it is held;
\item limiting access to PII to those individuals who require such access to perform their duties, and limit the access those individuals have to only that PII which they require access to in order to perform their duties;
\item resolving risks and vulnerabilities that are discovered through privacy risk assessments and audit processes; and
\item subjecting the controls to periodic review and reassessment in an ongoing security risk management process.

\end{itemize}

\textbf{11. Privacy compliance} Adhering to the privacy compliance principle means: 

\begin{itemize}

\item verifying and demonstrating that the processing meets data protection and privacy safeguarding requirements by periodically conducting audits using internal auditors or trusted third-party auditors;
\item having appropriate internal controls and independent supervision mechanisms in place that assure compliance with relevant privacy law and with their security, data protection and privacy policies and procedures; and
\item developing and maintaining privacy risk assessments in order to evaluate whether program and service delivery initiatives involving PII processing comply with data protection and privacy requirements.

\end{itemize}


\subsubsection{\textbf{Privacy by Design Principles for Big Data by Ann Cavoukian and Jeff Jonas}} 
Cavoukian and Jonas (2010) has extended the Cavoukian’s privacy principle \cite{Cavoukian2010}  as follows \cite{Cavoukian2012a} \label{Appendix:Privacy by Design Principles for Big Data by  and Ann Cavoukian Jeff Jonas}: 

\textbf{01. Full Attribution} Every observation (record) needs to know from where it came and when. There cannot be merge/purge data survivorship processing whereby some observations or fields are discarded.

\textbf{02. Data Tethering} Adds, changes and deletes occurring in systems of record must be accounted for, in real time, in sub-seconds.

\textbf{03. Analytics on Anonymised Data} The ability to perform advanced analytics (including some fuzzy matching) over cryptographically altered data means organizations can anonymise more data before information sharing.

\textbf{04. Tamper-Resistant Audit Logs} Every user search should be logged in a tamper-resistant manner-even the database administrator should not be able to alter the evidence contained in this audit log.

\textbf{05. False Negative Favouring Methods} The capability to more strongly favour false negatives is of critical importance in systems that could be used to affect someone’s civil liberties.

\textbf{06. Self-Correcting False Positives} With every new data point presented, prior assertions are re-evaluated to ensure they are still correct, and if no longer correct, these earlier assertions can often be repaired in real time.

\textbf{07. Information Transfer Accounting} Every secondary transfer of data, whether to human eyeball or a tertiary system, can be recorded to allow stakeholders (e.g., data custodians or the consumers themselves) to understand how their data is flowing.


\subsubsection{\textbf{ Wright and Raab Privacy Principles} \cite{Wright2014}:} 
\label{Appendix:Wright and Raab Privacy Principles}

\textbf{01. Right to dignity}, i.e., freedom from infringements upon the person or his / her reputation.

\textbf{02. Right to be let alone} (privacy of the home, etc.).

\textbf{03. Right to anonymity}, including the right to express oneâĂŹs views anonymously.

\textbf{04. Right to autonomy}, to freedom of thought and action, without being surveilled.

\textbf{ 05.} Right to individuality and uniqueness of identity.

\textbf{06.} Right to assemble or associate with others without being surveilled.

\textbf{07.} Right to confidentiality and secrecy of communications.

\textbf{08.} Right to travel (in physical or cyber space) without being tracked.

\textbf{09.} People should not have to pay in order to exercise their rights of privacy (subject to any justifiable exceptions), nor be denied goods or services or offered them on a less preferential basis.


\subsubsection{\textbf{Fisk et al. (2015) Principles} \cite{Fisk2015}:}
 \label{Appendix:Fisk et al. Principles}

\textbf{01. Principle of Least Disclosure} Systems should strive to disclose as little to others as possible, while still sharing.

\textbf{02. Principle of Qualitative Evaluation} One must balance (subjectively) costs and benefits for privacy and progress.

\textbf{03. Principle of Forward Progress} Organizations must not become paralyzed by Least Disclosure and Qualitative Evaluation.


\subsection{Privacy Strategies}\label{Appendix:Privacy Strategies}


\subsubsection{\textbf{Privacy Strategies by Rost and Bock }}\label{Appendix: Privacy Strategies by Rost and Bock} Rost and Bock (2011) has suggested six privacy strategies \cite{Rost2011}:

01. Availability

02. Integrity

03. Confidentiality

04. Transparency

05. Unlinkability

06. Ability to intervene


\subsubsection{\textbf{Privacy Strategies by Hoepman}} 
Hoepman (2014) has proposed eight privacy design strategies \cite{Hoepman2014}: \label{Appendix: Privacy Strategies by Hoepman}

\textbf{01. Minimise} The most basic privacy design strategy is MINIMISE, which states that the amount of personal data that is processed should be restricted to the minimal amount possible.

\textbf{02. Hide} Any personal data, and their interrelationships, should be hidden from plain view.

\textbf{03. Separate} Personal data should be processed in a distributed fashion, in separate compartments whenever possible.

\textbf{04. Aggregate} Personal data should be processed at the highest level of aggregation and with the least possible detail in which it is (still) useful.

\textbf{05. Inform} Data subjects should be adequately informed whenever personal data is processed. 

\textbf{06. Control} Data subjects should be provided agency over the processing of their personal data.

\textbf{07. Enforce} A privacy policy compatible with legal requirements should be in place and should be enforced.

\textbf{08. Demonstrate} Be able to demonstrate compliance with the privacy policy and any applicable legal requirements


\subsection{Privacy Guidelines}\label{Appendix:Privacy Guidelines}


\subsubsection{\textbf{Privacy Guidelines by O'Leary (1995)} \cite{DiscoveryOECD}:} \label{Appendix:Privacy Guidelines by Oleary}

\textbf{01. Collection limitation} Data should be obtained lawfully and fairly, while some very sensitive data should not be held at all.

\textbf{02. Data quality} Data should be relevant to the stated purposes, accurate, complete and up-to- date: proper precautions should be taken to ensure this accuracy.

\textbf{03. Purpose specification} The purposes for which data will be used should be identified, and the data should be destroyed if it no longer serves the given purpose.

\textbf{04. Use limitation} Use of data for purposes other than specified is forbidden, except with the consent of the data subject or by authority of the law.

\textbf{05. Security safeguards} Agencies should establish procedures to guard against loss. corruption, destruction, or misuse of data.

\textbf{06. Openness} It must be possible to acquire information about the collection, storage, and use of personal data.

\textbf{07. Individual participation} The data subject has a right to access and challenge the data related to him or her.

\textbf{08. Accountability} A data controller should be accountable for complying with measures giving effect to all these principles.


\subsubsection{\textbf{Privacy Guidelines by Perera et al. (2019) \cite{Perera2019}}:}
\label{Appendix:Privacy Guidelines by Perera}

\textbf{01. Minimise data acquisition} This guideline suggests to minimise the amount of data collected or requested by an IoT platform or application. Minimisation includes: 
\begin{itemize}
\item     
 Minimising data types (e.g., energy consumption, water consumption, temperature)
\item 
Minimum duration (e.g., hours, days, weeks, months) 
\item 
Minimum frequency (e.g., sampling rate: one second, 30 seconds, minutes)
\item 
Minimum amount of data (e.g., kilobytes, megabytes, gigabytes)

\end{itemize}

\textbf{02. Minimise number of data sources} This guideline suggests to minimise the number of data sources used by an IoT platform or application. Depending on the task at hand, it may be required to collect data from different sources. Multiple data sources may hold pieces of information about an individual (e.g., An activity tracking service may hold an individual’s activity data while a hospital may hold his health records). Aggregation of data from multiple sources allow malicious parties to identify sensitive personal information of an individual that that could lead to privacy violations.

\textbf{03. Minimise raw data intake} Whenever possible, IoT applications should reduce the amount of raw1 data intake. Raw data could lead to secondary usage and privacy violation. Therefore, IoT platforms should consider converting (or transforming) raw data into secondary context data. For example, IoT applications can extract orientation (e.g. sitting, standing, walking) by processing accelerometer data and store only the results (i.e. secondary context) and delete the raw accelerometer data.

\textbf{04. Minimise knowledge discovery} This guideline suggests to minimise the amount of knowledge discovered within an IoT application. IoT applications should only discover the knowledge necessary to achieve their primary objectives. For example, if the objective is to recommend food plans, it should not attempt to infer users’ health status without their explicit permission.

\textbf{05. Minimise data storage} This guideline suggests to minimise the amount of data (i.e. primary or secondary) stored by an IoT application. Any piece of data that is not required to perform a certain task should be deleted. For example, raw data can be deleted once secondary contexts are derived. Further, personally identifiable data may be deleted without storing.

\textbf{06. Minimise data retention period} This guideline suggests to minimise the duration for which data is stored (i.e. avoid retaining data for longer than needed). Long retention periods provide more time for malicious parties to attempt accessing the data in unauthorized manner. Privacy risks are also increased because long retention periods could lead to unconsented secondary usage.

\textbf{07. Hidden data routing} In IoT, data is generated within sensing devices. The data analysis typically happens within cloud servers. Therefore, data is expected to travel between different types of computational nodes before arriving at the processing cloud servers. This type of routing could reveal user locations and usage from anyone conducting network surveillance or traffic analysis. To make it more difficult for Internet activities to be traced back to the users, this guideline suggests that IoT applications should support and employ anonymous routing mechanism (e.g., torproject.org).

\textbf{08. Data anonymisation} This guideline suggests to remove personally identifiable information before the data gets used by IoT applications so that the people described by the data remain anonymous. Removal of personally identifiable information reduces the risk of unintended disclosure and privacy violations.

\textbf{09. Encrypted data communication} This guideline suggests that different components in an IoT application should consider encrypted data communication wherever possible. Encrypted data communication would reduce the potential privacy risks due to unauthorised access during data transfer between components. There are multiple data communication approaches based on the components involved in an IoT application, namely, 1) device-to-device, 2) device- to-gateway, 3) device-to-cloud, and 4)gateway-to-cloud. Sensor data communication can be encrypted using symmetric encryption AES 256 in the application layer. Typically, device- to-device communications are encrypted at the link layer using special electronic hardware included in the radio modules. Gateway-to-cloud communication is typically secured through HTTPS using Secure Sockets Layer (SSL) or Transport Layer Security (TLS).

\textbf{10. Encrypted data processing} This guideline suggests to process data while encrypted. Encryption is the process of encoding data in such a way that only authorised parties can read it. However, sometimes, the party who is responsible for processing data should not be allowed to read data. In such circumstances, it is important to process data while they are in encrypted form. For example, homomorphic encryption is a form of encryption that allows computations to be carried out on cipher-text, thus generating an encrypted result which, when decrypted, matches the result of operations performed on the plain-text.

\textbf{11. Encrypted data storage} This guideline suggests that IoT applications should store data in encrypted form. Encrypted data storage reduces any privacy violations due to malicious attacks and unauthorised access. Data encryption can be applied in different levels from sensors to the cloud. Depending on the circumstances, data can be encrypted using both hardware and software technologies.

\textbf{12. Reduce data granularity} The granularity is the level of depth represented by the data. High granularity refers to atomic grade of detail and low granularity zooms out into a summary view of data. For example, dissemination of location can be considered as coarse- grained and full address can be considered as fine-grained. Therefore, releasing fine grained information always has more privacy risks than coarse-grained data as they contain more information. Data granularity has a direct impact on the quality of the data as well as the accuracy of the results produced by processing such data. IoT applications should request the minimum level of granularity that is required to perform their primary tasks. Higher level of granularity could lead to secondary data usage and eventually privacy violations.

\textbf{13. Query answering} This guideline suggests to release high-level answers to the query when dissemination without releasing raw data. For example, a sample query would be ‘how energy  efficient a particular household is?’ where the answer would be in 0-5 scale. Raw data can always lead to privacy violations due to secondary usage. One such implementation is openPDS/SafeAnswers where it allows users to collect, store, and give high level answers to the queries while protecting their privacy. 

\textbf{14. Repeated query blocking} This guideline goes hand-in-hand with the Query answering guideline. When answering queries, IoT applications need to make sure that they block any malicious attempts to discover knowledge that violates user privacy through repeated queries (e.g, analysing intersections of multiple results).

\textbf{15. Distributed data processing} This guideline suggests that an IoT application should process data in a distributed manner. Similar, approaches are widely used in traditional wireless sensor network domain. Distributed processing avoids centralised large-scale data gathering. As a result, it deters any unauthorised data access attempts.

\textbf{16. Distributed data storage} This guideline recommends storing data in a distributed manner. Distributed data storage reduces any privacy violation due to malicious attacks and unauthorised access. It also reduces privacy risks due to unconsented secondary knowledge discovery.

\textbf{17. Knowledge discovery based aggregation} Aggregation of information over groups of attributes or groups of individuals, restricts the amount of detail in the personal data that remains. This guideline suggests to discover knowledge though aggregation and replace raw data with discovered new knowledge. For example, ‘majority of people who visited the park on [particular date] were young students’ is an aggregated result that is sufficient (once collected over a time period) to perform further time series based sales performance analysis of a near-by shop. Exact timings of the crowd movements are not necessary to achieve this objective.

\textbf{18. Geography based aggregation} This guideline recommends to aggregate data using geographical boundaries. For example, a query would be ‘how many electric vehicles used in each city in UK’. The results to this query would be an aggregated number unique to the each city. It is not required to collect or store detailed about individual electric vehicle.

\textbf{19. Chain aggregation} This guideline suggests to perform aggregation on-the-go while moving data from one node to another. For example, if the query requires a count or average, it can be done without pulling all the data items to a centralised location. Data will be sent from one node to another until all the nodes get a chance to respond. Similar techniques are successfully used in wireless sensor networks. This type of technique reduces the amount of data gathered by a centralised node (e.g. cloud server). Further, such aggregation also eliminates raw data from the results by reducing the risk of secondary data usage.

\textbf{20. Time-Period based aggregation} This guideline suggests to aggregate data over time (e.g. days, week, months). Aggregation reduces the granularity of data and also reduces the secondary usage that could lead to privacy violations. For example, energy consumption of a given house can be acquired and represented in aggregated form as 160 kWh per month instead of gathering energy consumption on daily or hourly basis.

\textbf{21. Category based aggregation} Categorisation based aggregation approaches can be used to reduce the granularity of the raw data. For example, instead of using exact value (e.g. 160 kWh per month), energy consumption of a given house can be represented as 150-200 kWh per month. Time-Period based and category based aggregation can be combined together to reduce data granularity.

\textbf{22. Information Disclosure} This guideline suggests that data subjects should be adequately informed whenever data they own is acquired, processed, and disseminated. Inform can take place at any stage of the data life cycle. Further, inform can be broadly divided into two categories: pre-inform and post-inform. Pre-inform takes place before data enters to a given data life cycle phase. Post-inform takes place soon after data leaves a given data life cycle phase.

\begin{itemize}
    \item 
Consent and Data Acquisition: what is the purpose of the data acquisition?, What types of data are requested?, What is the level of granularity?, What are the rights of the data subjects?
 \item 
 Data Pre-Processing: what data will be taken into the platform?, what data will be thrown out?, what kind of pre-processing technique will be employed?, what are the purposes of pre-processing data?, what techniques will be used to protect user privacy?
  \item 
 Data Processing and Analysis: what type of data will be analysed?, what knowledge will be discovered?, what techniques will be used?.
  \item 
 Data Storage: what data items will be stored? how long they will be stored? what technologies are used to store data (e.g. encryption techniques)? is it centralised or distributed storage? will there be any back up processes?
  \item 
 Data Dissemination: with whom the data will be shared? what rights will receivers have? what rights will data subjects have?

\end{itemize}

\textbf{23. Control} This guideline recommends providing privacy control mechanisms for data subjects. Control mechanisms will allow data owners to manage data based on their preference. There are different aspects that the data owner may like to control. However, controlling is a time consuming task and not every data owner will have the expertise to make such decisions. Therefore, it is a software architect’s responsibility to carefully go through the following list of possibilities and determine what kind of controls are useful and relevant to data owners in a given IoT application context. Further, it is important to provide some kind of default set of options for data owners to choose from, specially in the cases where data subjects do not have sufficient knowledge. Some potential aspects that a data owner may like to control are 1) data granularity, 2) anonymisation technique, 3) data retention period, 4) data dissemination.

\textbf{24. Logging} This guideline suggests to log events during all phases. It allows both internal and external parties to examine what has happened in the past to make sure a given system has performed as promised. Logging could include but not limited to event traces, performance parameters, timestamps, sequences of operations performed over data, any human interventions. For example, a log may include the timestamps of data arrival, operations performed in order to anonymise data, aggregation techniques performed, and so on.

\textbf{25. Auditing} This guideline suggests to perform systematic and independent examinations of logs, procedures, processes, hardware and software specifications, and so on. The logs above could play a significant role in this processes. Non-disclosure agreements may be helpful to allow auditing some parts of the classified data analytics processes.

\textbf{26. Open Source} Making source code of an IoT application open allows any external party to review code. Such reviews can be used as a form of compliance demonstration. This allows external parties to examine the code bases to verify and determine whether a given application or platform has taken all measures to protect user privacy.

\textbf{27. Data Flow} Data flow diagrams (e.g. Data Flow Diagrams used by Unified Modelling Language) allow interested parties to understand how data flows within a given IoT application and how data is being treated. Therefore, DFDs can be used as a form of a compliance demonstration.

\textbf{28. Certification} In this context, certification refers to the confirmation of certain characteristics of an system and process. Typically, certifications are given by a neutral authority. Certification will add trustworthiness to IoT applications. TRUSTe (truste.com) Privacy Seal is one example, even though none of the existing certifications are explicitly designed to certify IoT applications.

\textbf{29. Standardisation} This guideline suggests to follow standard practices as a way to demonstrate privacy protection capabilities. Industry wide standards (e.g. AllJoyn allseenalliance.org) typically inherit security measures that would reduce some privacy risks as well. This refers to the process of implementing and developing technical standards. Standardisation can help to maximise compatibility, interoperability, safety, repeatability, or quality. Standardisation will help external parties to easily understand the inner workings of a given IoT application.

\textbf{30. Compliance} Based on the country and region, there are number of policies, laws and regulations that need to be adhered to. It is important for IoT applications to respect guidelines. Some regulatory efforts are ISO 29100 Privacy framework, OECD privacy principles, and European Commission Protection of personal data.


\subsection{Privacy Patterns}\label{Appendix:Privacy Patterns}

This section presents the existing privacy patterns which are gathered from \cite{PrivacyPattern1,PrivacyPattern2}.

1. Protection against Tracking

2. Location Granularity

3. Minimal Information Asymmetry

4. Informed Secure Passwords

5. Awareness Feed

6. Encryption with user-managed keys

7. Federated Privacy Impact Assessment

8. Use of dummies

9. Who’s Listening

10. Privacy Policy Display

11. Layered Policy Design

12. Discouraging blanket strategies

13. Reciprocity

14. Asynchronous notice

15. Abridged Terms and Conditions

16. Policy Matching Display

17. Incentivized Participation

18. Outsourcing [with consent]

19. Ambient Notice

20. Dynamic Privacy Policy Display

21. Privacy Labels

22. Data Breach Notification Pattern

23. Pseudonymous Messaging

24. Onion Routing

25. Strip Invisible Metadata

26. Pseudonymous Identity

27. Personal Data Store

28. Trust Evaluation of Services Sides

29. Aggregation Gateway

30. Privacy icons

31. Privacy-Aware Network Client

32. Sign an Agreement to Solve Lack of Trust on the Use of Private Data Context

33. Single Point of Contact

34. Informed Implicit Consent

35. Enable/Disable Functions

36. Privacy Colour Coding

37. Appropriate Privacy Icons

38. User data confinement pattern

39. Icons for Privacy Policies

40. Obtaining Explicit Consent

41. Privacy Mirrors

42. Appropriate Privacy Feedback

43. Impactful Information and Feedback

44. Decoupling [content] and location information visibility

45. Platform for Privacy Preferences

46. Selective Access control

47. Pay Back

48. Privacy dashboard

49. Preventing mistakes or reducing their impact

50. Obligation Management

51. Informed Credential Selection

52. Anonymous Reputation-based Blacklisting

53. Negotiation of Privacy Policy

54. Reasonable Level of Control

55. Masquerade

56. Buddy List

57. Privacy Awareness Panel

58. Lawful Consent

59. Privacy Aware Wording

60. Sticky Policies

61. Personal Data Table

62. Informed Consent for Web-based Transactions

63. Added-noise measurement Obfuscation

64. Increasing awareness of information aggregation

65. Attribute Based Credentials

66. Trustworthy Privacy Plug-in

67. [Support] Selective Disclosure

68. Private link

69. Anonymity Set

70. Active broadcast of presence

71. Unusual Activities

72. Identity Federation Do Not Track Pattern

73. Dynamic Location Granularity


\section{Mapping between the Combined Privacy Laws Framework and Privacy by Design Schemes}\label{Appendix:Mapping between the Combined Privacy Laws Framework and Privacy by Design Schemes}

\subsubsection{\textbf{Mapping between the Combined Privacy Laws Framework and Privacy by Design Principles}}\label{subsubsec:Mapping Principles of the Combined Privacy Laws Framework and Privacy by Design Principles}
In this section, the objective is to correlate the principles of the Privacy by Design (PbD) schemes with the key principles and individuals’ rights of the CPLF based on large similarities between the description of each principle and right of the CPLF, with the description of each principle, guideline, and strategy of the PbD schemes. According to Table \ref{PrinciplesCateCavoukianISO}, Cate \cite{Cate2006} has proposed five principles (Section \ref{Appendix:Privacy by Design Principles by Cate (2010)}), Cavoukian \cite{Cavoukian2010} has suggested seven privacy principles (Section \ref{Appendix:Privacy by Design Principles by Ann Cavoukian (2010)}), and ISO/IEC 29100 \cite{Iso/Iec270322012} has issued 11 principles (Section \ref{Appendix:ISO 29100 Privacy framework}). According to Table \ref{PrinciplesCavoukianWrightFisk}, Cavoukian and Jonas \cite{Cavoukian2012a} have proposed 7 principles (Section \ref{Appendix:Privacy by Design Principles for Big Data by  and Ann Cavoukian Jeff Jonas}), Wright and Raab \cite{Wright2014} have created nine principles (Section \ref{Appendix:Wright and Raab Privacy Principles}), and three principles have been issued by Fisk et al. \cite{Fisk2015} (Section \ref{Appendix:Fisk et al. Principles}).

\begingroup
\tiny
 
        \begin{longtable}{ >{\raggedright}m{6.5em} ? m{0.17cm} | m{0.17cm} | m{0.17cm} | m{0.17cm} | m{0.17cm} ? m{0.17cm} | m{0.17cm} | m{0.17cm} | m{0.17cm} | m{0.17cm} | m{0.17cm} | m{0.17cm} ? m{0.17cm} | m{0.17cm} | m{0.17cm} | m{0.17cm} | m{0.17cm} | m{0.17cm} | m{0.17cm} | m{0.17cm} | m{0.17cm} | m{0.17cm} | m{0.17cm} }
        
\multicolumn{24}{l}{\textit{\textcolor{gray}{ - - Begin of Table}}}\\[0.5cm]  \caption {Mapping between the Combined Privacy Laws Framework and Privacy by Design Principles (Cate 2011, Cavoukian 2009, and ISO/IEC 29100 2011) \small{\textit{(shortened form of the terminologies - see Section \ref{Appendix:Privacy by Design Schemes} for the full form of PbD principles and Section \ref{Appendix:Result} for the full form of the CPLF)}}}\label{PrinciplesCateCavoukianISO} \\

\scriptsize{Principles} & \multicolumn{5}{p{2.1cm}}{Principles by Cate (2011) \cite{Cate2006}} |& \multicolumn{7}{p{3.3cm}}{Principles by Cavoukian (2010) \cite{Cavoukian2010}}
|& \multicolumn{11}{p{5.1cm}}{Principles by ISO/IEC 29100 (2011) \cite{Iso/Iec270322012}} \\
\hline
\endfirsthead
 
\multicolumn{24}{l}{\textit{\textcolor{gray}{- - Continuation of Table}}}\\[0.5cm]
\scriptsize{Principles} & \multicolumn{5}{p{2.1cm}}{Principles by Cate (2011) \cite{Cate2006}} |& \multicolumn{7}{p{3.3cm}}{Principles by Cavoukian (2010) \cite{Cavoukian2010}}
|& \multicolumn{11}{p{5.1cm}}{Principles by ISO/IEC 29100 (2011) \cite{Iso/Iec270322012}}  \\ 
\toprule
\scriptsize \textbf{Privacy Laws/ Rights} & \rotatebox{90}{0.1 Notice / Awareness ...} & \rotatebox{90}{0.2 Choice / Consent} & \rotatebox{90}{0.3 Access / Participation} & \rotatebox{90}{0.4 Integrity / Security ...} & \rotatebox{90}{0.5 Enforcement / Redress...} & \rotatebox{90}{01. Proactive not Reactive; ...} & \rotatebox{90}{02. Privacy as the ...} & \rotatebox{90}{03. Privacy Embedded ...} & \rotatebox{90}{04. Full Functionality-...} & \rotatebox{90}{05. End-to-End ...} & \rotatebox{90}{06. Visibility and Transparency - ...} & \rotatebox{90}{07. Respect for User ...} & \rotatebox{90}{01. Consent and choice} & \rotatebox{90}{02. Purpose legitimacy and ...} & \rotatebox{90}{03. Collection limitation} & \rotatebox{90}{04. Data minimisation} & \rotatebox{90}{05. Use, retention, and ...} & \rotatebox{90}{06. Accuracy and quality} & \rotatebox{90}{07. Openness, transparency, ...} & \rotatebox{90}{08. Individual participation ...} & \rotatebox{90}{09. Accountability} & \rotatebox{90}{10. Information Security} & \rotatebox{90}{11. Privacy Compliance} \\ \toprule [0.32ex]
 \endhead
 
 \endfoot

 \multicolumn{24}{l}{\textit{\textcolor{gray}{End of Table}}}\\[0.5cm]
 \endlastfoot

\toprule[0.32ex]
        \scriptsize \textbf{Privacy Laws/ Rights} & \rotatebox{90}{0.1 Notice / Awareness ...} & \rotatebox{90}{0.2 Choice / Consent} & \rotatebox{90}{0.3 Access / Participation} & \rotatebox{90}{0.4 Integrity / Security ...} & \rotatebox{90}{0.5 Enforcement / Redress...} & \rotatebox{90}{01. Proactive not Reactive; ...} & \rotatebox{90}{02. Privacy as the ...} & \rotatebox{90}{03. Privacy Embedded ...} & \rotatebox{90}{04. Full Functionality-...} & \rotatebox{90}{05. End-to-End ...} & \rotatebox{90}{06. Visibility and Transparency - ...} & \rotatebox{90}{07. Respect for User ...} & \rotatebox{90}{01. Consent and choice} & \rotatebox{90}{02. Purpose legitimacy and ...} & \rotatebox{90}{03. Collection limitation} & \rotatebox{90}{04. Data minimisation} & \rotatebox{90}{05. Use, retention, and ...} & \rotatebox{90}{06. Accuracy and quality} & \rotatebox{90}{07. Openness, transparency, ...} & \rotatebox{90}{08. Individual participation ...} & \rotatebox{90}{09. Accountability} & \rotatebox{90}{10. Information Security} & \rotatebox{90}{11. Privacy Compliance} \\ \toprule
        \scriptsize{Transparency} &\cellcolor{greeny!25}\ding{51}&&&&&&&&&&\cellcolor{greeny!25} \ding{51}&&\cellcolor{greeny!25} \ding{51}&\cellcolor{greeny!25} \ding{51} &&&&&\cellcolor{greeny!25} \ding{51}&&&& \\ \toprule
        \scriptsize{Purpose Limitation} &&&&&&\cellcolor{greeny!25} \ding{51}&\cellcolor{greeny!25} \ding{51}&&&&&&&\cellcolor{greeny!25} \ding{51}&&&&&&&&& \\ \toprule
        \scriptsize{Limiting Use, Disclosure} &&&&\cellcolor{greeny!25} \ding{51}&&\cellcolor{greeny!25} \ding{51}&\cellcolor{greeny!25} \ding{51}&\cellcolor{greeny!25} \ding{51}&&&&&&&&\cellcolor{greeny!25} \ding{51}&\cellcolor{greeny!25} \ding{51}&&&&&& \\ \toprule
        \scriptsize{Data Minimisation} &&&&&&\cellcolor{greeny!25} \ding{51}&\cellcolor{greeny!25} \ding{51}&\cellcolor{greeny!25} \ding{51}&&&&&&&\cellcolor{greeny!25} \ding{51}&\cellcolor{greeny!25} \ding{51}&&&&&&& \\ \toprule
        \scriptsize{Consent} &&&&&&\cellcolor{greeny!25} \ding{51}&&\cellcolor{greeny!25} \ding{51}&&&&\cellcolor{greeny!25} \ding{51}&\cellcolor{greeny!25} \ding{51}&&&&&&&&&& \\ \toprule
        \scriptsize{Lawfulness of Processing} &&&&&&\cellcolor{greeny!25} \ding{51}&&\cellcolor{greeny!25} \ding{51}&&&&\cellcolor{greeny!25} \ding{51}&\cellcolor{greeny!25} \ding{51}&&&&&&&&&& \\ \toprule
        \scriptsize{Accuracy} &&&\cellcolor{greeny!25} \ding{51}&\cellcolor{greeny!25} \ding{51}&&\cellcolor{greeny!25} \ding{51}&&&&&&&&&&&&\cellcolor{greeny!25} \ding{51}&&\cellcolor{greeny!25} \ding{51}&&& \\ \toprule
        \scriptsize{Storage Limitation} &&&&&&\cellcolor{greeny!25} \ding{51}&\cellcolor{greeny!25} \ding{51}&\cellcolor{greeny!25} \ding{51}&&&&&&&&\cellcolor{greeny!25} \ding{51}&\cellcolor{greeny!25} \ding{51}&&&&&& \\ \toprule
        \scriptsize{Security} &&&&\cellcolor{greeny!25} \ding{51}&&\cellcolor{greeny!25} \ding{51}&\cellcolor{greeny!25} \ding{51}&\cellcolor{greeny!25} \ding{51}&&\cellcolor{greeny!25} \ding{51}&&&&&&&&&&&&\cellcolor{greeny!25} \ding{51}& \\ \toprule
        \scriptsize{Accountability} &&&&&\cellcolor{greeny!25} \ding{51}&\cellcolor{greeny!25} \ding{51}&\cellcolor{greeny!25} \ding{51}&\cellcolor{greeny!25} \ding{51}&\cellcolor{greeny!25} \ding{51}&&&&&&&&&&&\cellcolor{greeny!25} \ding{51}&\cellcolor{greeny!25} \ding{51}&&\cellcolor{greeny!25} \ding{51} \\ \toprule
        \scriptsize{Anonymity ...} &&&&&&\cellcolor{greeny!25} \ding{51}&\cellcolor{greeny!25} \ding{51}&\cellcolor{greeny!25} \ding{51}&&\cellcolor{greeny!25} \ding{51}&&&&&&\cellcolor{greeny!25} \ding{51}&&&&&&& \\ \toprule
        \scriptsize{Source} &&&&&&&&\cellcolor{greeny!25} \ding{51}&&&&&&&&&&&&&&& \\ \toprule
        \scriptsize{Cross-border Disclousure ..} &&&&&&\cellcolor{greeny!25} \ding{51}&\cellcolor{greeny!25} \ding{51}&&&&&&&&&&&&&&&& \\ \toprule
        \scriptsize{Dealing ... Data} &&&&&&\cellcolor{greeny!25} \ding{51}&\cellcolor{greeny!25} \ding{51}&&&&&&&&&&&&&&&& \\ \toprule
        \scriptsize{Adoption, .. identifier} &&&&&&&\cellcolor{greeny!25} \ding{51}&\cellcolor{greeny!25} \ding{51}&&&&&&&&&&&&&&& \\ \toprule[0.32ex]
        \scriptsize{.. Exercise Rights} &&&\cellcolor{greeny!25} \ding{51}&&&&&&&&&\cellcolor{greeny!25} \ding{51}&&&&&&&&\cellcolor{greeny!25} \ding{51}&\cellcolor{greeny!25} \ding{51}&& \\ \toprule
        \scriptsize{Right to be informed} &\cellcolor{greeny!25} \ding{51}&&&&&&&&&&\cellcolor{greeny!25} \ding{51}&\cellcolor{greeny!25} \ding{51}&\cellcolor{greeny!25} \ding{51}&\cellcolor{greeny!25} \ding{51}&&&&&\cellcolor{greeny!25} \ding{51}&&&& \\ \toprule
        \scriptsize{Right of .. Access} &&&\cellcolor{greeny!25} \ding{51}&&&&&&&&&\cellcolor{greeny!25} \ding{51}&&&&&&&&\cellcolor{greeny!25} \ding{51}&&& \\ \toprule
        \scriptsize{Right to Rectification} &&&\cellcolor{greeny!25} \ding{51}&&&\cellcolor{greeny!25} \ding{51}&&&&&&\cellcolor{greeny!25} \ding{51}&&&&&&&&\cellcolor{greeny!25} \ding{51}&&& \\ \toprule
        \scriptsize{Right to Erasure} &&&&&&&&&&&&\cellcolor{greeny!25} \ding{51}&&&&&&&&\cellcolor{greeny!25} \ding{51}&&& \\ \toprule
        \scriptsize{.. Restriction of Processing} &&&&&&\cellcolor{greeny!25} \ding{51}&&&&&&\cellcolor{greeny!25} \ding{51}&&&&&&&&&&& \\ \toprule
        \scriptsize{Right to Object} &&&&&&\cellcolor{greeny!25} \ding{51}&&&&&&\cellcolor{greeny!25} \ding{51}&&&&&&&&&&& \\ \toprule
        \scriptsize{.. Object to Marketing} &&&&&&\cellcolor{greeny!25} \ding{51}&&&&&&\cellcolor{greeny!25} \ding{51}&&&&&&&&&&& \\ \toprule
        \scriptsize{Right to Data Portability} &&&&&&&&&&&&\cellcolor{greeny!25} \ding{51}&&&&&&&&&&& \\ \toprule
        \scriptsize{.. Automated Decision ..} &&&&&&\cellcolor{greeny!25} \ding{51}&&&&&&\cellcolor{greeny!25} \ding{51}&&&&&&&&&&& \\ \toprule
        \scriptsize{.. Withdraw Consent} &&\cellcolor{greeny!25} \ding{51}&&&&\cellcolor{greeny!25} \ding{51}&&&&&&\cellcolor{greeny!25} \ding{51}&\cellcolor{greeny!25} \ding{51}&&&&&&&&&& \\ \toprule
        \scriptsize{Right to Complain} &&&&&&&&&&&&\cellcolor{greeny!25} \ding{51}&&&&&&&&&\cellcolor{greeny!25} \ding{51}&& \\ \toprule
        \scriptsize{.. Discriminated} &&&&&&&&&&&&\cellcolor{greeny!25} \ding{51}&&&&&&&&&&& \\ \toprule[0.32ex]
     
    \end{longtable}

\endgroup

Regarding the relationship between the CPLF and the Privacy by Design (PbD) principles of Cate \cite{Cate2006}, Cavoukian \cite{Cavoukian2010}, and ISO/IEC \cite{Iso/Iec270322012}, Table \ref{PrinciplesCateCavoukianISO} shows that most of the principles and rights of the CPLF are associated with many of these PbD principles. Interestingly, the principles of Cavoukian \cite{Cavoukian2010} tend to cover all the principles and rights of the CPLF compared to Cate \cite{Cate2006}, and ISO/IEC \cite{Iso/Iec270322012} principles. Cavoukian \cite{Cavoukian2010}, furthermore, takes into consideration the interests of the individual under the principle of – \textbf{Respect for User Privacy; Keep it User-Centric} – as this principle achieves all the rights of the CPLF. This could simply return to the reason that Cavoukian \cite{Cavoukian2010} who first developed the concept of Privacy by Design. In contrast, \textbf{Consent, Lawfulness of processing, Storage limitation, and Anonymity and pseudonymity} principles of the CPLF are not covered by the principles of Cate \cite{Cate2006}. In addition, neither Cate \cite{Cate2006} nor ISO/IEC \cite{Iso/Iec270322012} consider the following principles of the CPLF: \textbf{Source, Cross-Border Disclosure of Personal Information, Dealing with Unsolicited Personal Data, and Adoption Use or Disclosure of an Identifier}. With regard to the rights, while Cavoukian \cite{Cavoukian2010} covers all the rights of the CPLF, Cate \cite{Cate2006} and ISO/IEC \cite{Iso/Iec270322012} have missed most of these rights.

\clearpage
\begingroup
\tiny
        \begin{longtable}{ >{\raggedright}m{10em} ? m{0.21cm} | m{0.21cm} | m{0.21cm} | m{0.21cm} | m{0.21cm} | m{0.21cm} | m{0.21cm} ? m{0.21cm} | m{0.21cm} | m{0.21cm} | m{0.21cm} | m{0.21cm} | m{0.21cm} | m{0.21cm} | m{0.21cm} | m{0.21cm} ? m{0.21cm} | m{0.21cm} | m{0.21cm} }
        
\multicolumn{20}{l}{\textit{\textcolor{gray}{ - - Begin of Table}}}\\[0.5cm]
 \caption {Mapping between the Combined Privacy Laws Framework and Privacy by Design Principles (Cavoukian et al. 2012, Wright et al. 2014, and Fisk et al. 2015)  \small{\textit{(shortened form of the terminologies; see Section \ref{Appendix:Privacy by Design Schemes} for the full form of PbD principles and Section \ref{Appendix:Result} for the full form of the CPLF)}}} \label{PrinciplesCavoukianWrightFisk} \\
\scriptsize{Principles} & \multicolumn{7}{p{2.1cm}}{Principles by Cavoukian and Jonas (2012) \cite{Cavoukian2012a}} |& \multicolumn{9}{p{2cm}}{Principles by Wright and Raab (2014) \cite{Wright2014}} |& \multicolumn{3}{p{1cm}}{Principles by Fisk et al. (2015) \cite{Fisk2015}}  \\
\hline
\endfirsthead
 
\multicolumn{20}{l}{\textit{\textcolor{gray}{- - Continuation of Table}}}\\[0.5cm]
\scriptsize{Principles} & \multicolumn{7}{p{2.1cm}}{Principles by Cavoukian and Jonas (2012) \cite{Cavoukian2012a}} |& \multicolumn{9}{p{2cm}}{Principles by Wright and Raab (2014) \cite{Wright2014}} |& \multicolumn{3}{p{1cm}}{Principles by Fisk et al. (2015) \cite{Fisk2015}}  \\ 
\toprule
\scriptsize \textbf{Privacy Laws/ Rights} & \rotatebox{90}{1. Full Attribution} & \rotatebox{90}{2. Data Tethering} & \rotatebox{90}{3. Analytics on Anonymised ...} & \rotatebox{90}{4. Tamper-Resistant ...} & \rotatebox{90}{5. False Negative ...} & \rotatebox{90}{6. Self-Correcting False ...} & \rotatebox{90}{7. Information Transfer ...} & \rotatebox{90}{01. Right to dignity} & \rotatebox{90}{02. Right to be let alone} & \rotatebox{90}{03. Right to anonymity} & \rotatebox{90}{04. Right to autonomy} & \rotatebox{90}{05. Right to individuality ...} & \rotatebox{90}{06. Right to assemble or associate ...} & \rotatebox{90}{07. Right to confidentiality and ...} & \rotatebox{90}{08. Right to travel ...} & \rotatebox{90}{09. People should not ...} & \rotatebox{90}{Principle of Least ...} & \rotatebox{90}{Principle of Qualitative ...} & \rotatebox{90}{Principle of Forward Progress} \\ \toprule [0.32ex]
 \endhead
 
 \endfoot

\multicolumn{20}{l}{\textit{\textcolor{gray}{End of Table}}}\\[0.5cm] \endlastfoot

\toprule[0.32ex]
\scriptsize \textbf{Privacy Laws/ Rights} & \rotatebox{90}{1. Full Attribution} & \rotatebox{90}{2. Data Tethering} & \rotatebox{90}{3. Analytics on Anonymised ...} & \rotatebox{90}{4. Tamper-Resistant ...} & \rotatebox{90}{5. False Negative ...} & \rotatebox{90}{6. Self-Correcting False ...} & \rotatebox{90}{7. Information Transfer ...} & \rotatebox{90}{01. Right to dignity} & \rotatebox{90}{02. Right to be let alone} & \rotatebox{90}{03. Right to anonymity} & \rotatebox{90}{04. Right to autonomy} & \rotatebox{90}{05. Right to individuality ...} & \rotatebox{90}{06. Right to assemble or associate ...} & \rotatebox{90}{07. Right to confidentiality and ...} & \rotatebox{90}{08. Right to travel ...} & \rotatebox{90}{09. People should not ...} & \rotatebox{90}{Principle of Least ...} & \rotatebox{90}{Principle of Qualitative ...} & \rotatebox{90}{Principle of Forward Progress} \\ \toprule
        \scriptsize{Transparency} &&&&&&&\cellcolor{greeny!25} \ding{51}&&&&&&&&\cellcolor{greeny!25} \ding{51}&&&& \\ \toprule
        \scriptsize{Purpose Limitation} &&&&&&&&&&&&&&&&&&& \\ \toprule
        \scriptsize{Limiting Use and Disclosure} &&&&&&&&&&&&&&&\cellcolor{greeny!25} \ding{51}&&\cellcolor{greeny!25} \ding{51}&&\cellcolor{greeny!25} \ding{51} \\ \toprule
        \scriptsize{Data Minimisation} &&&&&\cellcolor{greeny!25} \ding{51}&&&&\cellcolor{greeny!25} \ding{51}&\cellcolor{greeny!25} \ding{51}&\cellcolor{greeny!25} \ding{51}&\cellcolor{greeny!25} \ding{51}&\cellcolor{greeny!25} \ding{51}&&&&&& \\ \toprule
        \scriptsize{Consent} &&&&&&&&\cellcolor{greeny!25} \ding{51}&\cellcolor{greeny!25} \ding{51}&&\cellcolor{greeny!25} \ding{51}&\cellcolor{greeny!25} \ding{51}&&\cellcolor{greeny!25} \ding{51}&\cellcolor{greeny!25} \ding{51}&&&& \\ \toprule
        \scriptsize{Lawfulness of Processing} &&&&&&&&\cellcolor{greeny!25} \ding{51}&\cellcolor{greeny!25} \ding{51}&&\cellcolor{greeny!25} \ding{51}&\cellcolor{greeny!25} \ding{51}&&\cellcolor{greeny!25} \ding{51}&\cellcolor{greeny!25} \ding{51}&&&& \\ \toprule
        \scriptsize{Accuracy} &\cellcolor{greeny!25} \ding{51}&\cellcolor{greeny!25} \ding{51}&&&&\cellcolor{greeny!25} \ding{51}&&&&&&&&&&&&& \\ \toprule
        \scriptsize{Storage Limitation} &&&&&&&&&&&&&&&&&\cellcolor{greeny!25} \ding{51}&& \\ \toprule
        \scriptsize{Security} &&&\cellcolor{greeny!25} \ding{51}&\cellcolor{greeny!25} \ding{51}&&&&&&&&&&\cellcolor{greeny!25} \ding{51}&&&&& \\ \toprule
        \scriptsize{Accountability} &&&&&&&\cellcolor{greeny!25} \ding{51}&&&&&&&&&&&\cellcolor{greeny!25} \ding{51}&\cellcolor{greeny!25} \ding{51} \\ \toprule
        \scriptsize{Anonymity and Pseudonymity} &&&\cellcolor{greeny!25} \ding{51}&&&&&&\cellcolor{greeny!25} \ding{51}&\cellcolor{greeny!25} \ding{51}&\cellcolor{greeny!25} \ding{51}&\cellcolor{greeny!25} \ding{51}&\cellcolor{greeny!25} \ding{51}&&&&&& \\ \toprule
        \scriptsize{Source} &&&&&&&&&&&&&&&&&&& \\ \toprule
        \scriptsize{Cross-border Disclousure ...} &&&&&&&&&&&&&&&&&&& \\ \toprule
        \scriptsize{Dealing with Unsolicited Data} &&&&&&&&&&&&&&&&&&& \\ \toprule
        \scriptsize{Adoption ... an identifier} &&&&&&&&&&&&&&&&&&& \\ \toprule[0.32ex]
        \scriptsize{Right ... Exercise their Rights} &&&&&&&&&&&&&&&&\cellcolor{greeny!25} \ding{51}&&& \\ \toprule
        \scriptsize{Right to be informed} &&&&&&&\cellcolor{greeny!25} \ding{51}&&&&&&&&&&&& \\ \toprule
        \scriptsize{Right of Individuals Access} &&&&&&&&&&&&&&&&&&& \\ \toprule
        \scriptsize{Right to Rectification} &\cellcolor{greeny!25} \ding{51}&\cellcolor{greeny!25} \ding{51}&&&&\cellcolor{greeny!25} \ding{51}&&&&&&&&&&&&& \\ \toprule
        \scriptsize{Right to Erasure} &&&&&&&&&&&&&&&&&&& \\ \toprule
        \scriptsize{Right to Restriction of Processing} &&&&&&&&&&&&&&&&&&& \\ \toprule
        \scriptsize{Right to Object} &&&&&&&&&&&&&&&\cellcolor{greeny!25} \ding{51}&&&& \\ \toprule
        \scriptsize{Right to Object to Marketing} &&&&&&&&&&&&&&&&&&& \\ \toprule
        \scriptsize{Right to Data Portability} &&&&&&&&&&&&&&&&&&& \\ \toprule
        \scriptsize{Right to object to Automated ...} &&&&&&&&&&&&&&&&&&& \\ \toprule
        \scriptsize{Right to Withdraw Consent} &&&&&&&&&&&&&&&&&&& \\ \toprule
        \scriptsize{Right to Complain} &&&&&&&&&&&&&&&&&&& \\ \toprule
        \scriptsize{Right ... not to  be Discriminated} &&&&&&&&&&&&&&&&\cellcolor{greeny!25} \ding{51}&&& \\
        \toprule[0.32ex]

    \end{longtable}

\endgroup

According to Table \ref{PrinciplesCavoukianWrightFisk}, it seems that the principles of Cavoukian and Joans \cite{Cavoukian2012a}, Wright and Raab \cite{Wright2014}, and Fisk et al. \cite{Fisk2015} do not cover many of the CPLF's principles. This could be due to the nature of these principles, as the principles of Cavoukian and Joans are concerned with directing technical outcomes. While these principles move consumer privacy issues away from a policy or compliance issue towards a business imperative \cite{Cavoukian2012a}, the principles and rights of the CPLF tend to be legal requirements rather than technical requirements. As a result, only some – \textbf{Transparency, Accuracy, Security, Accountability} and \textbf{Anonymity and pseudonymity} – of the CPLF principles are achieved. \textbf{Right to be Informed}, and \textbf{Right to Rectification} – are also covered. With regard to the principles of Wright and Raab \cite{Wright2014}, they do not aim at creating principles that cover all the privacy issues, but aim to identify further privacy principles for the existing privacy principles that could be applied to different types of privacy with consideration of the risks and harms of different types of privacy policies \cite{Wright2014}. Accordingly, seven out of 15 CPLF principles have been achieved, as shown in Table 4, and only two of the CPLF rights are covered – \textbf{Right of Individuals to Exercise their Rights}, and \textbf{Right to Object}. Fisk et al. \cite{Fisk2015}, in contrast, define three principles that focus on sharing security information between organisations \cite{Fisk2015}. This results in achieving only three of the principles – \textbf{Limiting Use and Disclosure, Storage Limitation}, and \textbf{Accountability} – of the CPLF and missing all of its rights.


\subsubsection{\textbf{Mapping between the Combined Privacy Laws Framework and Privacy by Design Strategies}}\label{subsubsec:Mapping between the Combined Privacy Laws Framework and Privacy by Design Strategies}
 This section aims at correlating the strategies of Privacy by Design (PbD) and the key principles and individuals’ rights of the CPLF. According to Table \ref{StrategiesRostHoepman}, while six privacy strategies have been issued by Rost and Bock \cite{Rost2011} (Section \ref{Appendix: Privacy Strategies by Rost and Bock}), Hoepman \cite{Hoepman2014} has created eight privacy strategies (see Section \ref{Appendix: Privacy Strategies by Hoepman}).

\begingroup
\tiny
\begin{longtable}{ >{\raggedright}m{18em} ? m{0.3cm} | m{0.3cm} | m{0.3cm} | m{0.3cm} | m{0.3cm} | m{0.3cm} ? m{0.3cm} | m{0.3cm} | m{0.3cm} | m{0.3cm} | m{0.3cm} | m{0.3cm} | m{0.3cm} | m{0.3cm} }

\multicolumn{15}{l}{\textit{\textcolor{gray}{ - - Begin of Table}}}\\[0.5cm]
 \caption{Mapping between Privacy Laws and Privacy by Design Strategies (Rost et al. 2011, and Hoepman 2014) \small{\textit{((shorten form of the terminologies; see Section \ref{Appendix:Result} for the full form of the CPLF)}}}  \label{StrategiesRostHoepman}    \\
\scriptsize{Strategies } & \multicolumn{6}{p{2.8cm}}{Strategies by Rost and Bock (2011) } \cite{Rost2011} |& \multicolumn{8}{p{2.6cm}}{Strategies by Hoepman (2014)} \cite{Hoepman2014}
\endfirsthead
 
\multicolumn{15}{l}{\textit{\textcolor{gray}{- - Continuation of Table}}}\\[0.5cm]
\scriptsize{Strategies } & \multicolumn{6}{p{2.8cm}}{Strategies by Rost and Bock (2011) \cite{Rost2011}} |& \multicolumn{8}{p{2.6cm}}{Strategies by Hoepman (2014) \cite{Hoepman2014}}  \\ 
\toprule
\scriptsize \textbf{Continuation of Privacy Laws/ Rights} & \rotatebox{90}{01. Availability} & \rotatebox{90}{02. Integrity} & \rotatebox{90}{03. Confidentiality} & \rotatebox{90}{04. Transparency} & \rotatebox{90}{05. Unlinkability} & \rotatebox{90}{06. Ability to intervene} & \rotatebox{90}{01. Minimise} & \rotatebox{90}{02. Hide} & \rotatebox{90}{03. Separate} & \rotatebox{90}{04. Aggregate} & \rotatebox{90}{05. Inform} & \rotatebox{90}{06. Control} & \rotatebox{90}{07. Enforce} & \rotatebox{90}{08. Demonstrate} \\ \toprule [0.32ex]
 \endhead
 
 \endfoot

\multicolumn{15}{l}{\textit{\textcolor{gray}{End of Table}}}\\[0.5cm]

 \endlastfoot

\toprule[0.32ex]
\scriptsize \textbf{Privacy Laws/ Rights} & \rotatebox{90}{01. Availability} & \rotatebox{90}{02. Integrity} & \rotatebox{90}{03. Confidentiality} & \rotatebox{90}{04. Transparency} & \rotatebox{90}{05. Unlinkability} & \rotatebox{90}{06. Ability to intervene} & \rotatebox{90}{01. Minimise} & \rotatebox{90}{02. Hide} & \rotatebox{90}{03. Separate} & \rotatebox{90}{04. Aggregate} & \rotatebox{90}{05. Inform} & \rotatebox{90}{06. Control} & \rotatebox{90}{07. Enforce} & \rotatebox{90}{08. Demonstrate} \\ \toprule
        \scriptsize{Transparency} &&&&\cellcolor{greeny!25} \ding{51}&&&&&&&\cellcolor{greeny!25} \ding{51}&&& \\ \toprule
        \scriptsize{Purpose Limitation} &&&&&\cellcolor{greeny!25} \ding{51}&&&&&&&&\cellcolor{greeny!25} \ding{51}& \\ \toprule
        \scriptsize{Limiting Use and Disclosure} &&&&&&&\cellcolor{greeny!25} \ding{51}&&&&&&& \\ \toprule
        \scriptsize{Data Minimisation} &&&&&&&\cellcolor{greeny!25} \ding{51}&\cellcolor{greeny!25} \ding{51}&&\cellcolor{greeny!25} \ding{51}&&&& \\ \toprule
        \scriptsize{Consent} &&&&&&\cellcolor{greeny!25} \ding{51}&&&&&&\cellcolor{greeny!25} \ding{51}&& \\ \toprule
        \scriptsize{Lawfulness of Processing} &&&&&&\cellcolor{greeny!25} \ding{51}&&&&&&\cellcolor{greeny!25} \ding{51}&& \\ \toprule
        \scriptsize{Accuracy} &&\cellcolor{greeny!25} \ding{51}&&&&&&&&&&&& \\ \toprule
        \scriptsize{Storage Limitation} &&&&&&&\cellcolor{greeny!25} \ding{51}&&&&&&& \\ \toprule
        \scriptsize{Security} &&&\cellcolor{greeny!25} \ding{51}&&&&&\cellcolor{greeny!25} \ding{51}&\cellcolor{greeny!25} \ding{51}&\cellcolor{greeny!25} \ding{51}&&& \\ \toprule
        \scriptsize{Accountability} &\cellcolor{greeny!25} \ding{51}&&&\cellcolor{greeny!25} \ding{51}&\cellcolor{greeny!25} \ding{51} &\cellcolor{greeny!25} \ding{51}&&&&&&&\cellcolor{greeny!25} \ding{51}&\cellcolor{greeny!25} \ding{51} \\ \toprule
        \scriptsize{Anonymity and Pseudonymity} &&&&&&&\cellcolor{greeny!25} \ding{51}&\cellcolor{greeny!25} \ding{51}&&&&&& \\ \toprule
        \scriptsize{Source} &&&&&&&&&&&&&& \\ \toprule
        \scriptsize{Cross-border Disclousure ...} &&&&&&&&&&&&&& \\ \toprule
        \scriptsize{Dealing with Unsolicited Data} &&&&&&&&&&&&&& \\ \toprule
        \scriptsize{Adoption, Use ... of an Identifier} &&&&&&&&&&&&&& \\ \toprule[0.32ex]
        \scriptsize{Right ... to exercise their Rights} &&&&&&\cellcolor{greeny!25} \ding{51}&&&&&&\cellcolor{greeny!25} \ding{51}&& \\ \toprule
        \scriptsize{Right to be informed} &&&&&&&&&&&\cellcolor{greeny!25} \ding{51}&&& \\ \toprule
        \scriptsize{Right of Individuals Access} &&&&&&\cellcolor{greeny!25} \ding{51}&&&&&&\cellcolor{greeny!25} \ding{51}&& \\ \toprule
        \scriptsize{Right to Rectification} &&&&&&\cellcolor{greeny!25} \ding{51}&&&&&&\cellcolor{greeny!25} \ding{51}&& \\ \toprule
        \scriptsize{Right to Erasure} &&&&&&\cellcolor{greeny!25} \ding{51}&&&&&&\cellcolor{greeny!25} \ding{51}&& \\ \toprule
        \scriptsize{Right to Restriction of Processing} &&&&&&\cellcolor{greeny!25} \ding{51}&&&&&&\cellcolor{greeny!25} \ding{51}&& \\ \toprule
        \scriptsize{Right to Object} &&&&&&\cellcolor{greeny!25} \ding{51}&&&&&&\cellcolor{greeny!25} \ding{51}&& \\ \toprule
        \scriptsize{Right to Object to Marketing} &&&&&&\cellcolor{greeny!25} \ding{51}&&&&&&\cellcolor{greeny!25} \ding{51}&& \\ \toprule
        \scriptsize{Right to Data Portability} &&&&&&\cellcolor{greeny!25} \ding{51}&&&&&&\cellcolor{greeny!25} \ding{51}&& \\ \toprule
        \scriptsize{Right to object to Automated Decision-Making} &&&&&&\cellcolor{greeny!25} \ding{51}&&&&&&\cellcolor{greeny!25} \ding{51}&& \\ \toprule
        \scriptsize{Right to Withdraw Consent} &&&&&&\cellcolor{greeny!25} \ding{51}&&&&&&\cellcolor{greeny!25} \ding{51}&& \\ \toprule
        \scriptsize{Right to Complain} &&&&&&&&&&&&&& \\ \toprule
        \scriptsize{Right ... not to  be Discriminated}&&&&&&&&&&&&&& \\ \toprule[0.32ex]
    \end{longtable}
\endgroup

According to Table \ref{StrategiesRostHoepman}, Rost and Bock \cite{Rost2011} have formulated new data protection goals which are classified in this paper under the strategy of PbD schemes as it has an equivalent meaning, where strategy achieves a certain design goal. Hoepman \cite{Hoepman2014} also defines eight PbD strategies from the perspective of an IT system, and considers legal requirements to be the point of departure. Accordingly, most of the principles and rights of the CPLF are covered by these strategies. More precisely, as shown in Table \ref{StrategiesRostHoepman}, these PbD strategies tend to cover more of the rights of the CPLF compared to the principles of CPLF. However, both strategies do not take into consideration the following principles of the CPLF: – \textbf{Source, Cross-border Disclosure of Personal Data, Dealing with Unsolicited Personal Data,} and \textbf{Adoption, Use, Disclosure of an Identifier}. While three strategies of Hoepman \cite{Hoepman2014} cover the principle of – \textbf{Data Minimisation} – , none of Rost and Bock's \cite{Rost2011} strategies are concerned about the \textbf{Data Minimisation} principle. In addition, neither the – \textbf{Limiting Use and Disclosure} – nor – \textbf{Storage Limitation} – and – \textbf{Anonymity and Pseudonymity} – of the CPLF are achieved by Rost and Bock's \cite{Rost2011} strategies. This could simply refer to the perspective of Rost and Bock \cite{Rost2011} as they claim that their strategies do not stand alone but should be combined together with the privacy by design principles to be a comprehensive and universally accepted concept \cite{Rost2011}. By contrast, while Rost's and Bock's \cite{Rost2011} strategies achieve the \textbf{Accuracy} principle, Hoepman’s \cite{Hoepman2014} strategies do not.


\subsubsection{\textbf{Mapping between the Combined Privacy Laws Framework and Privacy by Design Guidelines}}\label{subsubsec:Mapping Principles of the Combined Privacy Laws Framework and Privacy by Design Guidelines}
In this section, the guidelines by O’Leary \cite{DiscoveryOECD} (Section \ref{Appendix:Privacy Guidelines by Oleary}) the guidelines by Perera et al.\cite{Perera2019} (see Section \ref{Appendix:Privacy Guidelines by Perera}) are mapped to the key principles and individuals’ rights in the CPLF, as demonstrated in Tables \ref{GuidlinesOleary} and \ref{tab:GuidelinesPerera}.

\begingroup
\tiny
        \begin{longtable}{ >{\raggedright}m{24em} ? >{\centering}m{0.7cm} | >{\centering}m{0.7cm} | >{\centering}m{0.7cm} | >{\centering}m{0.7cm} | >{\centering}m{0.7cm} | >{\centering}m{0.7cm} | >{\centering}m{0.7cm} | m{0.7cm}}

\multicolumn{9}{l}{\textit{\textcolor{gray}{ - - Begin of Table}}}\\[0.5cm] \caption{Mapping between the Combined Privacy Laws Framework and Privacy by Design Guidelines
(O’Leary 1995) \cite{DiscoveryOECD} \small{\textit{(shortened form of the terminologies; see Section \ref{Appendix:Result} for the full form of the CPLF)}}}\label{GuidlinesOleary}\\

\scriptsize{Guidelines } & \multicolumn{8}{p{4.4cm}}{Guidelines by O'Leary (1995) \cite{DiscoveryOECD}} \\
\hline
\endfirsthead
 
\multicolumn{9}{l}{\textit{\textcolor{gray}{- - Continuation of Table}}}\\[0.5cm]
\scriptsize{Guidelines } & \multicolumn{8}{p{4.4cm}}{Guidelines by O'Leary (1995) \cite{DiscoveryOECD}}  \\ 
\toprule
\scriptsize \textbf{Continuation of Privacy Laws/ Rights} & \rotatebox{90}{01. Collection limitation} & \rotatebox{90}{02. Data quality} & \rotatebox{90}{03. Purpose specification} & \rotatebox{90}{04. Use limitation} & \rotatebox{90}{05. Security safeguards} & \rotatebox{90}{06. Openness} & \rotatebox{90}{07. Individual participation} & \rotatebox{90}{08. Accountability} \\ \toprule [0.32ex]
 \endhead
 
 \endfoot

 \multicolumn{9}{l}{\textit{\textcolor{gray}{End of Table}}}\\[0.5cm]

 \endlastfoot

\toprule[0.32ex]
\scriptsize \textbf{Privacy Laws/ Rights} & \rotatebox{90}{01. Collection limitation} & \rotatebox{90}{02. Data quality} & \rotatebox{90}{03. Purpose specification} & \rotatebox{90}{04. Use limitation} & \rotatebox{90}{05. Security safeguards} & \rotatebox{90}{06. Openness} & \rotatebox{90}{07. Individual participation} & \rotatebox{90}{08. Accountability} \\ \toprule
        \scriptsize{Transparency} &&&&&&\cellcolor{greeny!25} \ding{51}&& \\ \toprule
        \scriptsize{Purpose Limitation} &&&\cellcolor{greeny!25} \ding{51}&&&&& \\ \toprule
        \scriptsize{Limiting Use and Disclosure} &&&&\cellcolor{greeny!25} \ding{51}&&&& \\ \toprule
        \scriptsize{Data Minimisation} &\cellcolor{greeny!25} \ding{51}&&&&&&& \\ \toprule
        \scriptsize{Consent} &&&&&&&& \\ \toprule
        \scriptsize{Lawfulness of Processing} &\cellcolor{greeny!25} \ding{51}&&&&&&& \\ \toprule
        \scriptsize{Accuracy} &&\cellcolor{greeny!25} \ding{51}&&&&&& \\ \toprule
        \scriptsize{Storage Limitation} &&&\cellcolor{greeny!25} \ding{51}&&&&& \\ \toprule
        \scriptsize{Security} &&&&&\cellcolor{greeny!25} \ding{51}&&& \\ \toprule
        \scriptsize{Accountability} &&&&&&&&\cellcolor{greeny!25} \ding{51} \\ \toprule
        \scriptsize{Anonymity and Pseudonymity} &&&&&&&& \\ \toprule
        \scriptsize{Source} &&&&&&&& \\ \toprule
        \scriptsize{Cross-border Disclousure of Personal Infomation} &&&&&&&& \\ \toprule
        \scriptsize{Dealing with Unsolicited Data} &&&&&&&& \\ \toprule
        \scriptsize{Adoption, Use or Disclosure of an identifier} &&&&&&&& \\ \toprule[0.32ex]
        \scriptsize{Right of individuals to exercise their rights} &&&&&&&\cellcolor{greeny!25} \ding{51}& \\ \toprule
        \scriptsize{Right to be informed} &&&&&&\cellcolor{greeny!25} \ding{51}&& \\ \toprule
        \scriptsize{Right of Individuals Access} &&&&&&&\cellcolor{greeny!25} \ding{51}& \\ \toprule
        \scriptsize{Right to Rectification} &&&&&&&\cellcolor{greeny!25} \ding{51}& \\ \toprule
        \scriptsize{Right to Erasure} &&&&&&&\cellcolor{greeny!25} \ding{51}& \\ \toprule
        \scriptsize{Right to Restriction of Processing} &&&&&&&\cellcolor{greeny!25} \ding{51}& \\ \toprule
        \scriptsize{Right to Object} &&&&&&&\cellcolor{greeny!25} \ding{51}& \\ \toprule
        \scriptsize{Right to Object to Marketing} &&&&&&&\cellcolor{greeny!25} \ding{51}& \\ \toprule
        \scriptsize{Right to Data Portability} &&&&&&&& \\ \toprule
        \scriptsize{Right to object to Automated Decision-Making} &&&&&&&\cellcolor{greeny!25} \ding{51}& \\ \toprule
        \scriptsize{Right to Withdraw Consent} &&&&&&&& \\ \toprule
        \scriptsize{Right to Complain} &&&&&&&& \\  \toprule
         \scriptsize{Right ... not to  be Discriminated} &&&&&&&& \\ \toprule[0.32ex]

    \end{longtable}
\endgroup

With regard to the correlation of the CPLF and PbD guidelines by O’Leary \cite{DiscoveryOECD}, Table \ref{GuidlinesOleary} shows that O’Leary’s guidelines cover most of the principles and rights of the CPLF. Considering the right of the CPLF, as seen in Table \ref{GuidlinesOleary}, all the rights of the CPLF are achieved by O’Leary \cite{DiscoveryOECD} except the – \textbf{Right to Data Portability}, \textbf{Right to Withdraw Consent}, \textbf{Right to Complain}, and \textbf{Right of Individuals not to  be Discriminated}. According to the principles of the CPLF, O’Leary’s \cite{DiscoveryOECD} guidelines cover many of these principles: –\textbf{Transparency}, \textbf{Purpose Limitation}, \textbf{Limiting Use and Disclosure}, \textbf{Data Minimisation}, \textbf{Lawfulness of processing}, \textbf{Accuracy}, \textbf{Storage Limitation}, \textbf{Security}, and \textbf{Accountability}.


\begingroup
\tiny
\begin{longtable}{ >{\raggedright}m{5.5em} ? m{0.045cm} | m{0.045cm} | m{0.045cm} | m{0.045cm} | m{0.045cm} | m{0.045cm} | m{0.045cm} | m{0.045cm} | m{0.045cm} | m{0.045cm} | m{0.045cm} | m{0.045cm} | m{0.045cm} | m{0.045cm} | m{0.045cm} | m{0.045cm} | m{0.045cm} | m{0.045cm} | m{0.045cm} | m{0.045cm} | m{0.045cm} | m{0.045cm} | m{0.045cm} | m{0.045cm} | m{0.045cm} | m{0.045cm} | m{0.045cm} | m{0.045cm} | m{0.045cm} | m{0.045cm}}
        
\multicolumn{31}{l}{\textit{\textcolor{gray}{ - - Begin of Table}}}\\[0.5cm]  \caption{Mapping between the Combined Privacy Laws Framework and  Privacy by Design Guidelines (Perera et al. 2019) \small{\textit{(shortened form of the terminologies; see Section \ref{Appendix:Result} for the full form of the CPLF)}}}  \label{tab:GuidelinesPerera} \\

\scriptsize{Guidelines } & \multicolumn{30}{p{3cm}}{Guidelines by Perera et al. (2019) \cite{Perera2019}} \\
\hline
\endfirsthead
 
\multicolumn{31}{l}{\textit{\textcolor{gray}{- - Continuation of Table}}}\\[0.5cm]
\scriptsize{Guidelines } & \multicolumn{30}{p{3cm}}{Guidelines by Perera et al. (2019) \cite{Perera2019}} \\ 
\toprule
\scriptsize \textbf{Privacy Laws/ Rights} & \rotatebox{90}{01. Minimise data acquisition} & \rotatebox{90}{02. Minimise number of ...} & \rotatebox{90}{03. Minimise raw data ...} & \rotatebox{90}{04. Minimise knowledge ...} & \rotatebox{90}{05. Minimise data storage} & \rotatebox{90}{06. Minimise data retention ...} & \rotatebox{90}{07. Hidden data routing} & \rotatebox{90}{08. Data anonymisation} & \rotatebox{90}{09. Encrypted data communication} & \rotatebox{90}{10. Encrypted data processing} & \rotatebox{90}{11. Encrypted data storage} & \rotatebox{90}{12. Reduce data granularity} & \rotatebox{90}{13. Query answering} & \rotatebox{90}{14. Repeated query blocking} & \rotatebox{90}{15. Distributed data processing} & \rotatebox{90}{16. Distributed data storage} & \rotatebox{90}{17. Knowledge discovery based ...} & \rotatebox{90}{18. Geography based aggregation} & \rotatebox{90}{19. Chain aggregation} & \rotatebox{90}{20. Time-Period based ...} & \rotatebox{90}{21. Category based aggregation} & \rotatebox{90}{22. Information Disclosure} & \rotatebox{90}{23. Control} & \rotatebox{90}{24. Logging} & \rotatebox{90}{25. Auditing} & \rotatebox{90}{26. Open Source} & \rotatebox{90}{27. Data Flow} & \rotatebox{90}{28. Certification} & \rotatebox{90}{29. Standardisation} & \rotatebox{90}{30. Compliance} \\ \toprule [0.32ex]
 \endhead
 
 \endfoot

 \multicolumn{31}{l}{\textit{\textcolor{gray}{End of Table}}}\\[0.5cm]

 \endlastfoot

\toprule[0.32ex]
\scriptsize \textbf{Privacy Laws/ Rights} & \rotatebox{90}{01. Minimise data acquisition} & \rotatebox{90}{02. Minimise number of ...} & \rotatebox{90}{03. Minimise raw data ...} & \rotatebox{90}{04. Minimise knowledge ...} & \rotatebox{90}{05. Minimise data storage} & \rotatebox{90}{06. Minimise data retention ...} & \rotatebox{90}{07. Hidden data routing} & \rotatebox{90}{08. Data anonymisation} & \rotatebox{90}{09. Encrypted data communication} & \rotatebox{90}{10. Encrypted data processing} & \rotatebox{90}{11. Encrypted data storage} & \rotatebox{90}{12. Reduce data granularity} & \rotatebox{90}{13. Query answering} & \rotatebox{90}{14. Repeated query blocking} & \rotatebox{90}{15. Distributed data processing} & \rotatebox{90}{16. Distributed data storage} & \rotatebox{90}{17. Knowledge discovery based ...} & \rotatebox{90}{18. Geography based aggregation} & \rotatebox{90}{19. Chain aggregation} & \rotatebox{90}{20. Time-Period based ...} & \rotatebox{90}{21. Category based aggregation} & \rotatebox{90}{22. Information Disclosure} & \rotatebox{90}{23. Control} & \rotatebox{90}{24. Logging} & \rotatebox{90}{25. Auditing} & \rotatebox{90}{26. Open Source} & \rotatebox{90}{27. Data Flow} & \rotatebox{90}{28. Certification} & \rotatebox{90}{29. Standardisation} & \rotatebox{90}{30. Compliance} \\ \toprule
        \scriptsize{Transparency} &&&&& &&&&& &&&&& &&&&& &&&&& &&&&& \\ \toprule
        \scriptsize{Purpose Limitation} &&&&\cellcolor{greeny!25} \ding{51}& &&&&& &&&&& &&&&& &&&&& &&&&& \\ \toprule
        \scriptsize{Limiting Use ..} &&&\cellcolor{greeny!25} \ding{51}&\cellcolor{greeny!25} \ding{51}& &&&&& &&\cellcolor{greeny!25} \ding{51}&\cellcolor{greeny!25} \ding{51} && &&\cellcolor{greeny!25} \ding{51}&&\cellcolor{greeny!25} \ding{51}&\cellcolor{greeny!25} \ding{51} &\cellcolor{greeny!25} \ding{51}&&&& &&&&& \\ \toprule
        \scriptsize{Data Minimisation} &\cellcolor{greeny!25} \ding{51}&\cellcolor{greeny!25} \ding{51}&\cellcolor{greeny!25} \ding{51}&&\cellcolor{greeny!25} \ding{51} &&&&& &&\cellcolor{greeny!25} \ding{51}&\cellcolor{greeny!25} \ding{51}&& &&\cellcolor{greeny!25} \ding{51}&\cellcolor{greeny!25} \ding{51}&\cellcolor{greeny!25} \ding{51}& &\cellcolor{greeny!25} \ding{51}&&&& &&&&& \\ \toprule
        \scriptsize{Consent} &&&&& &&&&& &&&&& &&&&& &&&&& &&&&& \\ \toprule
        \scriptsize{Lawfulness of Processing} &&&&& &&&&& &&&&& &&&&& &&&&& &&&&& \\ \toprule
        \scriptsize{Accuracy} &&&&& &&&&& &&&&& &&&&& &&&&& &&&&& \\ \toprule
        \scriptsize{Storage Limitation} &&&\cellcolor{greeny!25} \ding{51}&& &\cellcolor{greeny!25} \ding{51}&&&& &&&&& &&&&& &&&&& &&&&& \\ \toprule
        \scriptsize{Security} &&&&& &&\cellcolor{greeny!25} \ding{51}& &\cellcolor{greeny!25} \ding{51}&\cellcolor{greeny!25} \ding{51} &\cellcolor{greeny!25} \ding{51}&&&\cellcolor{greeny!25} \ding{51}&\cellcolor{greeny!25} \ding{51} &\cellcolor{greeny!25} \ding{51}&&&\cellcolor{greeny!25} \ding{51}& &&&&& &&&&& \\ \toprule
        \scriptsize{Accountability} &&&&& &&&&& &&&&& &&&&& &&&&\cellcolor{greeny!25} \ding{51}&\cellcolor{greeny!25} \ding{51} &\cellcolor{greeny!25} \ding{51}&\cellcolor{greeny!25} \ding{51}&\cellcolor{greeny!25} \ding{51}&\cellcolor{greeny!25} \ding{51}&\cellcolor{greeny!25} \ding{51} \\ \toprule
        \scriptsize{Anonymity and Pseudonymity} &&&&& &&&\cellcolor{greeny!25} \ding{51}&& &&&&& &&&&& &&&&& &&&&& \\ \toprule
        \scriptsize{Source} &&&&& &&&&& &&&&& &&&&& &&&&& &&&&& \\ \toprule
        \scriptsize{Cross-border Disclosure ..} &&&&& &&&&& &&&&& &&&&& &&&&& &&&&& \\ \toprule
        \scriptsize{Dealing with Unsolicited Data} &&&&& &&&&& &&&&& &&&&& &&&&& &&&&& \\ \toprule
        \scriptsize{Adoption ... an identifier} &&&&& &&&&& &&&&& &&&&& &&&&& &&&&& \\ \toprule[0.32ex]
        \scriptsize{Right ... to exercise their rights} &&&&& &&&&& &&&&& &&&&& &&&\cellcolor{greeny!25} \ding{51}&& &&&&& \\ \toprule
        \scriptsize{Right to be informed} &&&&& &&&&& &&&&& &&&&& &&\cellcolor{greeny!25} \ding{51}&&& &&&&& \\ \toprule
        \scriptsize{Right of Individuals Access} &&&&& &&&&& &&&&& &&&&& &&&\cellcolor{greeny!25} \ding{51}&& &&&&& \\ \toprule
        \scriptsize{Right to Rectification} &&&&& &&&&& &&&&& &&&&& &&&\cellcolor{greeny!25} \ding{51}&& &&&&& \\ \toprule
        \scriptsize{Right to Erasure} &&&&& &&&&& &&&&& &&&&& &&&\cellcolor{greeny!25} \ding{51}&& &&&&& \\ \toprule
        \scriptsize{Right to Restriction ...} &&&&& &&&&& &&&&& &&&&& &&&\cellcolor{greeny!25} \ding{51}&& &&&&& \\ \toprule
        \scriptsize{Right to Object} &&&&& &&&&& &&&&& &&&&& &&&\cellcolor{greeny!25} \ding{51}&& &&&&& \\ \toprule
        \scriptsize{Right to Object to Marketing} &&&&& &&&&& &&&&& &&&&& &&&\cellcolor{greeny!25} \ding{51}&& &&&&& \\ \toprule
        \scriptsize{Right to Data Portability} &&&&& &&&&& &&&&& &&&&& &&&\cellcolor{greeny!25} \ding{51}&& &&&&& \\ \toprule
        \scriptsize{Right to ... Automated ...} &&&&& &&&&& &&&&& &&&&& &&&\cellcolor{greeny!25} \ding{51}&& &&&&& \\ \toprule
        \scriptsize{Right to Withdraw Consent} &&&&& &&&&& &&&&& &&&&& &&&&& &&&&& \\ \toprule
        \scriptsize{Right to Complain}&&&&& &&&&& &&&&& &&&&& &&&&& &&&&& \\ \toprule
        \scriptsize{Right .. Discriminated} &&&&&&&&&&&&&&&&&&&&&&&&&&&&&  \\ \toprule[0.32ex]
  
    \end{longtable}
\endgroup

Table \ref{tab:GuidelinesPerera} shows the correlation between the Privacy by Design (PbD) guidelines by Perera et al. \cite{Perera2019} and the CPLF. These 30 guidelines are inspired by Hoepman’s strategies and have been developed for the IoT domain specifically to help software engineers develop and assess IoT applications \cite{Perera2019}. As seen in Table \ref{tab:GuidelinesPerera}, various principles and rights of the CPLF are achieved by Perera et al.’s \cite{Perera2019} guidelines. Nevertheless, these guidelines tend to focus on achieving some of the CPLF’s principles through several of the guidelines, rather than covering all the principles of the CPLF. For example, the – \textbf{Data Minimisation} – principle of the CPLF is covered by ten of Perera et al.’s \cite{Perera2019} guidelines, while the – \textbf{Transparency}– principle is not covered by any of these guidelines. With regards to the rights of the CPLF, interestingly, all of these rights are covered by Perera et al.’s \cite{Perera2019} guidelines except, \textbf{Right to Withdraw Consent}, \textbf{Right to Complain}, and \textbf{Right of Individuals not to  be Discriminated}.


\subsection{Mapping Privacy Laws and Privacy Patterns}\label{subsec:Mapping Privacy Laws and Privacy Patterns}

This section correlates the privacy patterns and the principles and rights of the CPLF. These privacy patterns are gathered from \cite{PrivacyPattern1,PrivacyPattern2}, as many common patterns are shared between these two resources. The principles and the rights of the CPLF are usually aimed at the legal domain and not the technical domain due to remaining disconnected from developers’ tools and the environment required for building applications. Therefore, mapping the principles and the rights of the CPLF with the privacy patterns is essential to facilitate developers building a privacy-aware application and complying with the law. Typically, patterns are more reliable and it is possible to explain their use in a particular context, compared to guidelines, strategies or principles for PbD schemes, which is highlighted in Table \ref{PrinciplesCateCavoukianISO}, Table \ref{PrinciplesCavoukianWrightFisk}, Table \ref{StrategiesRostHoepman}, Table \ref{GuidlinesOleary}, and Table \ref{tab:GuidelinesPerera}. The following criteria are followed in order to map between the privacy laws and the privacy patterns:\\

\begingroup
\tiny
\begin{tabu}to \textwidth {X[l]}
\toprule
$\bullet$ \hspace{0.2cm} Direct relationship: If a specific pattern can achieve a law (principle/right) directly (a direct law). \\ \hline
$\circ$ \hspace{0.2cm} Indirect relationship: If a specific law can achieve a specific pattern through a direct law. \\ \hline
\toprule
\end{tabu}
\endgroup

\begingroup
\renewcommand\arraystretch{0.74}
\tiny
\begin{longtable}{ >{\raggedright}m{3.2cm} ? m{0.03cm} | m{0.03cm} | m{0.03cm} | m{0.03cm} | m{0.03cm} | m{0.03cm} | m{0.03cm} | m{0.03cm} | m{0.03cm} | m{0.03cm} | m{0.03cm} | m{0.03cm} | m{0.03cm} | m{0.03cm} | m{0.03cm} ? m{0.03cm} | m{0.03cm} | m{0.03cm} | m{0.03cm} | m{0.03cm} | m{0.03cm} | m{0.03cm} | m{0.03cm} | m{0.03cm} | m{0.03cm} | m{0.03cm} | m{0.03cm} | m{0.03cm} }
\caption {Mapping between the Combined Privacy Laws Framework and Privacy by Design Patterns \small{\textit{(shortened form of the terminologies; see Section \ref{Appendix:Privacy Patterns} for the full form of privacy patterns and Section \ref{Appendix:Result} for the full form of the CPLF)}}} \label{tab:Mapping between the Combined Privacy Laws Framework and Privacy by Design Patterns}\\
       
 \multicolumn{24}{l}{\textit{\textcolor{gray}{ - - Begin of Table}}}\\[0.5cm]
Privacy Patterns & \rotatebox{90}{Transparency} & \rotatebox{90}{Purpose Limitation} & \rotatebox{90}{Limiting Use and Disclosure} & \rotatebox{90}{Data Minimisation} & \rotatebox{90}{Consent} & \rotatebox{90}{Lawfulness of Processing} & \rotatebox{90}{Accuracy} & \rotatebox{90}{Storage Limitation} & \rotatebox{90}{Security} & \rotatebox{90}{Accountability} & \rotatebox{90}{Anonymity and Pseudonymity} & \rotatebox{90}{Source} & \rotatebox{90}{Cross-border Disclosure ...} & \rotatebox{90}{Dealing with Unsolicited Data} & \rotatebox{90}{Adoption ... an identifier} & \rotatebox{90}{Right ... to exercise their rights} & \rotatebox{90}{Right to be informed} & \rotatebox{90}{Right of Individuals Access} & \rotatebox{90}{Right to Rectification} & \rotatebox{90}{Right to Erasure} & \rotatebox{90}{Right to Restriction ...} & \rotatebox{90}{Right to Object} & \rotatebox{90}{Right to Object to Marketing} & \rotatebox{90}{Right to Data Portability} & \rotatebox{90}{Right to object to Automated ...} & \rotatebox{90}{Right to Withdraw Consent} & \rotatebox{90}{Right to Complain} & \rotatebox{90}{Right ... not to  be Discriminated}  \\
\hline
\endfirsthead
  
\multicolumn{24}{l}{\textit{\textcolor{gray}{- - Continuation of Table}}}\\[0.5cm]
Privacy Patterns & \rotatebox{90}{Transparency} & \rotatebox{90}{Purpose Limitation} & \rotatebox{90}{Limiting Use and Disclosure} & \rotatebox{90}{Data minimisation} & \rotatebox{90}{Consent} & \rotatebox{90}{Lawfulness of Processing} & \rotatebox{90}{Accuracy} & \rotatebox{90}{Storage Limitation} & \rotatebox{90}{Security} & \rotatebox{90}{Accountability} & \rotatebox{90}{Anonymity and Pseudonymity} & \rotatebox{90}{Source} & \rotatebox{90}{Cross-border Disclosure ...} & \rotatebox{90}{Dealing with Unsolicited Data} & \rotatebox{90}{Adoption ... an identifier} & \rotatebox{90}{Right ... to exercise their rights} & \rotatebox{90}{Right to be informed} & \rotatebox{90}{Right of Individuals Access} & \rotatebox{90}{Right to Rectification} & \rotatebox{90}{Right to Erasure} & \rotatebox{90}{Right to Restriction ...} & \rotatebox{90}{Right to Object} & \rotatebox{90}{Right to Object to Marketing} & \rotatebox{90}{Right to Data Portability} & \rotatebox{90}{Right to object to Automated ...} & \rotatebox{90}{Right to Withdraw Consent} & \rotatebox{90}{Right to Complain} & \rotatebox{90}{Right ... not to  be Discriminated}   \\
\toprule[0.32ex]
 \endhead
 
 \endfoot
 
 \multicolumn{24}{l}{\textit{\textcolor{gray}{End of Table}}}\\[0.5cm]

 \endlastfoot

\toprule[0.32ex]
        \scriptsize{1. Protection against ...}                         &&\cellcolor{yellowy!25} $\circ$&&& &&&\cellcolor{greeny!25} $\bullet$&\cellcolor{yellowy!25} $\circ$&\cellcolor{yellowy!25} $\circ$ &\cellcolor{greeny!25} $\bullet$&&&& &&&&& &&&& &&& \\ \toprule
        \scriptsize{2. Location Granularity}                                &&\cellcolor{yellowy!25} $\circ$&\cellcolor{greeny!25} $\bullet$&\cellcolor{greeny!25} $\bullet$ & & & & & &\cellcolor{yellowy!25} $\circ$ & & & & & & & & & & & & & & & & &  \\ \toprule
        \scriptsize{3. Minimal Information ...}                       &\cellcolor{greeny!25} $\bullet$&\cellcolor{yellowy!25} $\circ$&&\cellcolor{greeny!25} $\bullet$& &&&&&\cellcolor{yellowy!25} $\circ$ &&&&& &&\cellcolor{greeny!25} $\bullet$&&& &&&& &&& \\ \toprule
        \scriptsize{4. Informed Secure Passwords}                           &&&&& &&&&\cellcolor{greeny!25} $\bullet$&\cellcolor{yellowy!25} $\circ$ &&&&& &&&&& &&&& &&& \\ \toprule
        \scriptsize{5. Awareness Feed}                                      &\cellcolor{greeny!25} $\bullet$&&&& &&&&&\cellcolor{yellowy!25} $\circ$ &&&&& &&\cellcolor{greeny!25} $\bullet$&&& &&&& &&& \\ \toprule
        \scriptsize{6. Encryption ...}                   &&&&& &&&&\cellcolor{greeny!25} $\bullet$&\cellcolor{yellowy!25} $\circ$ &&&&& &&&&& &&&& &&& \\ \toprule
        \scriptsize{7. Federated Privacy ...}                 &&&&& &&&&&\cellcolor{greeny!25} $\bullet$ &&&&& &&&&& &&&& &&& \\ \toprule
        \scriptsize{8. Use of Dummies}                                      &&&&& &&&&\cellcolor{yellowy!25} $\circ$&\cellcolor{yellowy!25} $\circ$ &\cellcolor{greeny!25} $\bullet$&&&& &&&&& &&&& &&& \\ \toprule
        \scriptsize{9. Who’s Listening}                                     &&&&& &&&&&\cellcolor{yellowy!25} $\circ$ &&&&& &&\cellcolor{greeny!25} $\bullet$&&& &&&& &&& \\ \toprule
        \scriptsize{10. Privacy Policy Display}                              &\cellcolor{greeny!25} $\bullet$&&&& &&&&&\cellcolor{yellowy!25} $\circ$ &&&&& &&\cellcolor{greeny!25} $\bullet$&&& &&&& &&& \\ \toprule
        \scriptsize{11. Layered Policy Design}                               &\cellcolor{greeny!25} $\bullet$&&&& &&&&&\cellcolor{yellowy!25} $\circ$ &&&&& &&\cellcolor{greeny!25} $\bullet$&&& &&&& &&& \\ \toprule
        \scriptsize{12. Discouraging Blanket ...}                     &&&&& &&&&& &&&&& &&&&& &&&& &&& \\ \toprule
        \scriptsize{13. Reciporcity}                                         &&&&& &&&&& &&&&& &&&&& &&&& &&& \\ \toprule
        \scriptsize{14. Asynchronous Notice}                                 &&&&& &&&&&\cellcolor{yellowy!25} $\circ$ &&&&& &&\cellcolor{greeny!25} $\bullet$&&& &&&& &&& \\ \toprule
        \scriptsize{15. Abridged Terms ...}                        &\cellcolor{greeny!25} $\bullet$&&&& &&&&&\cellcolor{yellowy!25} $\circ$ &&&&& &&\cellcolor{greeny!25} $\bullet$&&& &&&& &&& \\ \toprule
        \scriptsize{16. Policy Matching Display}                             &&&&& &&&&& &&&&& &&&&& &&&& &&& \\ \toprule
        \scriptsize{17. Incentivized Participation}                          &&&&& &&&&& &&&&& &&&&& &&&& &&& \\ \toprule
        \scriptsize{18. Outsourcing [with consent]}                          &&\cellcolor{yellowy!25} $\circ$&\cellcolor{greeny!25} $\bullet$&&\cellcolor{greeny!25} $\bullet$ &\cellcolor{greeny!25} $\bullet$&&&&\cellcolor{yellowy!25} $\circ$ &&&&& &&&&& &&&& &&& \\ \toprule
        \scriptsize{19. Ambient Notice}                                      &\cellcolor{greeny!25} $\bullet$&&&& &&&&&\cellcolor{yellowy!25} $\circ$ &&&&& &&\cellcolor{greeny!25} $\bullet$&&& &&&& &&& \\ \toprule
        \scriptsize{20. Dynamic Privacy Policy ...}                      &\cellcolor{greeny!25} $\bullet$&&&& &&&&&\cellcolor{yellowy!25} $\circ$ &&&&& &&\cellcolor{greeny!25} $\bullet$&&& &&&& &&& \\ \toprule
        \scriptsize{21. Privacy Labels}                                      &\cellcolor{greeny!25} $\bullet$&&&& &&&&&\cellcolor{yellowy!25} $\circ$ &&&&& &&\cellcolor{greeny!25} $\bullet$&&& &&&& &&& \\ \toprule
        \scriptsize{22. Data Breach Notification ...}                    &&&&& &&&&\cellcolor{greeny!25} $\bullet$&\cellcolor{greeny!25} $\bullet$ &&&&& &&&&& &&&& &&& \\ \toprule
        \scriptsize{23. Pseudonymous Messaging}                              &&&&& &&&&\cellcolor{greeny!25} $\bullet$&\cellcolor{yellowy!25} $\circ$ &\cellcolor{greeny!25} $\bullet$&&&& &&&&& &&&& &&& \\ \toprule
        \scriptsize{24. Onion Routing}                                       &&&&& &&&&\cellcolor{greeny!25} $\bullet$&\cellcolor{yellowy!25} $\circ$ &\cellcolor{greeny!25} $\bullet$&&&& &&&&& &&&& &&& \\ \toprule
        \scriptsize{25. Strip Invisible Metadata}                            &&\cellcolor{yellowy!25} $\circ$&\cellcolor{greeny!25} $\bullet$&\cellcolor{greeny!25} $\bullet$& &&&&&\cellcolor{yellowy!25} $\circ$ &&&&& &&&&& &&&& &&& \\ \toprule
        \scriptsize{26. Pseudonymous Identitiy}                              &&&&& &&&&\cellcolor{yellowy!25} $\circ$&\cellcolor{yellowy!25} $\circ$ &\cellcolor{greeny!25} $\bullet$&&&& &&&&& &&&& &&& \\ \toprule
        \scriptsize{27. Personal Data Store}                                 &&&&& &&&&& &&&&& &&&&& &&&& &&& \\ \toprule
        \scriptsize{28. Trust Evaluation ...}                 &\cellcolor{greeny!25} $\bullet$&&&& &&&&&\cellcolor{greeny!25} $\bullet$ &&&&& &&\cellcolor{greeny!25} $\bullet$&&& &&&& &&& \\ \toprule
        \scriptsize{29. Aggregation Gateway}                                 &&&&\cellcolor{yellowy!25} $\circ$& &&&&\cellcolor{greeny!25} $\bullet$&\cellcolor{yellowy!25} $\circ$ &&&&& &&&&& &&&& &&& \\ \toprule
        \scriptsize{30. Privacy icons}                                       &\cellcolor{greeny!25} $\bullet$&&&& &&&&&\cellcolor{yellowy!25} $\circ$ &&&&& &&\cellcolor{greeny!25} $\bullet$&&& &&&& &&& \\ \toprule
        \scriptsize{31. Privacy-Aware Network ...} &\cellcolor{greeny!25} $\bullet$&&&& &&&&&\cellcolor{yellowy!25} $\circ$ &&&&& &&\cellcolor{greeny!25} $\bullet$&&& &&&& &&& \\ \toprule
        \scriptsize{32. Sign an Agreement ...} &&&&&\cellcolor{greeny!25} $\bullet$ &\cellcolor{greeny!25} $\bullet$&&&&\cellcolor{yellowy!25} $\circ$ &&&&& &&&&& &&&& &&& \\ \toprule
        \scriptsize{33. Single Point of ...} &&&&& &&&&\cellcolor{greeny!25} $\bullet$&\cellcolor{yellowy!25} $\circ$ &&&&& &&&&& &&&& &&& \\ \toprule
        \scriptsize{34. Informed Implicit ...} &\cellcolor{greeny!25} $\bullet$&&&&\cellcolor{greeny!25} $\bullet$ &\cellcolor{yellowy!25} $\circ$&&&&\cellcolor{yellowy!25} $\circ$ &&&&& &&\cellcolor{greeny!25} $\bullet$&&& &&&& &&& \\ \toprule
        \scriptsize{35. Enable/Disable Function} &\cellcolor{greeny!25} $\bullet$&&&&\cellcolor{greeny!25} $\bullet$ &\cellcolor{yellowy!25} $\circ$&&&&\cellcolor{yellowy!25} $\circ$ &&&&& &&\cellcolor{greeny!25} $\bullet$&&& &&&& &&& \\ \toprule
        \scriptsize{36. Privacy Color Coding} &\cellcolor{greeny!25} $\bullet$&&&& &&&&&\cellcolor{yellowy!25} $\circ$ &&&&& &&\cellcolor{greeny!25} $\bullet$&&& &&&& &&& \\ \toprule
        \scriptsize{37. Appropriate Privacy Icons} &\cellcolor{greeny!25} $\bullet$&&&& &&&&&\cellcolor{yellowy!25} $\circ$ &&&&& &&\cellcolor{greeny!25} $\bullet$&&& &&&& &&& \\ \toprule
        \scriptsize{38. User Data Confinement ... } &&\cellcolor{yellowy!25} $\circ$&&\cellcolor{greeny!25} $\bullet$& &&&&&\cellcolor{yellowy!25} $\circ$ &&&&& &&&&& &&&& &&& \\ \toprule
        \scriptsize{39. Icons for Privacy Policies} &\cellcolor{greeny!25} $\bullet$&&&& &&&&&\cellcolor{yellowy!25} $\circ$ &&&&& &&\cellcolor{greeny!25} $\bullet$&&& &&&& &&& \\ \toprule
        \scriptsize{40. Obtaining Explicit Consent} &&&&&\cellcolor{greeny!25} $\bullet$ &\cellcolor{greeny!25} $\bullet$&&&&\cellcolor{yellowy!25} $\circ$ &&&&& &&&&& &&&& &&& \\ \toprule
        \scriptsize{41. Privacy Mirrors} &&&&& &&&&&\cellcolor{yellowy!25} $\circ$ &&&&& &&\cellcolor{greeny!25} $\bullet$&\cellcolor{greeny!25} $\bullet$&& &&&& &&& \\ \toprule
        \scriptsize{42. Appropriate Privacy ...} &\cellcolor{greeny!25} $\bullet$&&&& &&&&&\cellcolor{yellowy!25} $\circ$ &&&&& &&\cellcolor{greeny!25} $\bullet$&&& &&&& &&& \\ \toprule
        \scriptsize{43. Impactful Information ...} &&&&& &&&&&\cellcolor{yellowy!25} $\circ$ &&&&& &&\cellcolor{greeny!25} $\bullet$&&& &&&& &&& \\ \toprule
        \scriptsize{44. Decoupling [content] ...} &&&&& &&&&\cellcolor{greeny!25} $\bullet$&\cellcolor{yellowy!25} $\circ$ &&&&& &&&&& &&&& &&& \\ \toprule
        \scriptsize{45. Platform for Privacy ...} &\cellcolor{greeny!25} $\bullet$&&&& &&&&&\cellcolor{yellowy!25} $\circ$ &&&&& &&\cellcolor{greeny!25} $\bullet$&&& &&&& &&& \\ \toprule
        \scriptsize{46. Selective Access ..,} &&&&& &&&&\cellcolor{greeny!25} $\bullet$&\cellcolor{yellowy!25} $\circ$ &&&&& &&&&& &&&& &&& \\ \toprule
        \scriptsize{47. Pay Back} &&&&& &&&&& &&&&& &&&&& &&&& &&& \\ \toprule
        \scriptsize{48. Privacy Dashboard} &\cellcolor{greeny!25} $\bullet$&&&& &&&&&\cellcolor{yellowy!25} $\circ$ &&&&& &\cellcolor{greeny!25} $\bullet$&\cellcolor{greeny!25} $\bullet$&\cellcolor{greeny!25} $\bullet$&\cellcolor{greeny!25} $\bullet$&\cellcolor{greeny!25} $\bullet$ &\cellcolor{greeny!25} $\bullet$&\cellcolor{greeny!25} $\bullet$&\cellcolor{greeny!25} $\bullet$& &\cellcolor{greeny!25} $\bullet$&\cellcolor{greeny!25} $\bullet$&& \\ \toprule
        \scriptsize{49. Preventing Mistakes ...} &&\cellcolor{yellowy!25} $\circ$&\cellcolor{greeny!25} $\bullet$&&\cellcolor{yellowy!25} $\circ$ &\cellcolor{yellowy!25} $\circ$&&&&\cellcolor{yellowy!25} $\circ$ &&&&& &&\cellcolor{greeny!25} $\bullet$&&& &&&&& &&& \\ \toprule
        \scriptsize{50. Obligation Management} &&\cellcolor{yellowy!25} $\circ$&&\cellcolor{greeny!25} $\bullet$& &&&\cellcolor{greeny!25} $\bullet$&&\cellcolor{greeny!25} $\bullet$ &&&&& &&\cellcolor{greeny!25} $\bullet$&&& &&&&& &&& \\ \toprule
        \scriptsize{51. Informed Credential ...} &\cellcolor{greeny!25} $\bullet$&\cellcolor{yellowy!25} $\circ$&&\cellcolor{greeny!25} $\bullet$& &&&&&\cellcolor{yellowy!25} $\circ$ &&&&& &&\cellcolor{greeny!25} $\bullet$&&& &&&&& &&& \\ \toprule
        \scriptsize{52. Anonymous Reputation ...} &&&&& &&&&\cellcolor{greeny!25} $\bullet$&\cellcolor{yellowy!25} $\circ$ &\cellcolor{greeny!25} $\bullet$&&&& &&&&& &&&&& &&& \\ \toprule
        \scriptsize{53. Negotiation of ...  Policy} &&&&& &&&&& &&&&& &&&&& &&&&& &&& \\ \toprule
        \scriptsize{54. Reasonable Level of ...} &&\cellcolor{yellowy!25} $\circ$&\cellcolor{greeny!25} $\bullet$&& &&&&&\cellcolor{yellowy!25} $\circ$ &&&&& &&&&& &&&&& &&& \\ \toprule
        \scriptsize{55. Masquerade} &&\cellcolor{yellowy!25} $\circ$&\cellcolor{greeny!25} $\bullet$&& &&&&&\cellcolor{yellowy!25} $\circ$ &&&&& &&&&& &&&&& &&& \\ \toprule
        \scriptsize{56. Buddy List} &&\cellcolor{yellowy!25} $\circ$&\cellcolor{greeny!25} $\bullet$&& &&&&&\cellcolor{yellowy!25} $\circ$ &&&&& &&&&& &&&&& &&& \\ \toprule
        \scriptsize{57. Privacy Awareness Panel} &\cellcolor{greeny!25} $\bullet$&&&& &&&&&\cellcolor{yellowy!25} $\circ$ &&&&& &&\cellcolor{greeny!25} $\bullet$&&& &&&&& &&& \\ \toprule
        \scriptsize{58. Lawful Consent} &&\cellcolor{greeny!25} $\bullet$&&&\cellcolor{greeny!25} $\bullet$ &\cellcolor{greeny!25} $\bullet$&&&&\cellcolor{yellowy!25} $\circ$ &&&&& &&&&& &&&&& &&& \\ \toprule
        \scriptsize{59. Privacy Aware Wording} &\cellcolor{greeny!25} $\bullet$&&&& &&&&&\cellcolor{yellowy!25} $\circ$ &&&&& &&\cellcolor{yellowy!25} $\circ$&&& &&&&& &&& \\ \toprule
        \scriptsize{60. Sticky Policies} &&\cellcolor{yellowy!25} $\circ$&&\cellcolor{greeny!25} $\bullet$& &&&\cellcolor{greeny!25} $\bullet$&&\cellcolor{greeny!25} $\bullet$ &&&&& &&\cellcolor{greeny!25} $\bullet$&&& &&&&& &&& \\ \toprule
        \scriptsize{61. Personal Data Table} &\cellcolor{greeny!25} $\bullet$&&&& &&&&&\cellcolor{yellowy!25} $\circ$ &&&&& &\cellcolor{greeny!25} $\bullet$&&\cellcolor{greeny!25} $\bullet$&& &&&&& &&& \\ \toprule
        \scriptsize{62. Informed Consent ...} &&&&&\cellcolor{greeny!25} $\bullet$ &\cellcolor{greeny!25} $\bullet$&&&&\cellcolor{yellowy!25} $\circ$ &&&&& &&&&& &&&&& &&& \\ \toprule
        \scriptsize{63. Added-noise ...} &&\cellcolor{yellowy!25} $\circ$&\cellcolor{greeny!25} $\bullet$&& &&&&&\cellcolor{yellowy!25} $\circ$ &&&&& &&&&& &&&&& &&& \\ \toprule
        \scriptsize{64. Increasing Awareness ...} &\cellcolor{greeny!25} $\bullet$&&&&\cellcolor{yellowy!25} $\circ$ &&&&&\cellcolor{yellowy!25} $\circ$ &&&&& &&\cellcolor{greeny!25} $\bullet$&&& &&&&& &&& \\ \toprule
        \scriptsize{65. Attribute Based ...} &&\cellcolor{yellowy!25} $\circ$&&\cellcolor{greeny!25} $\bullet$& &&&&\cellcolor{yellowy!25} $\circ$&\cellcolor{yellowy!25} $\circ$ &\cellcolor{greeny!25} $\bullet$&&&& &&&&& &&&&& &&& \\ \toprule
        \scriptsize{66. Trustworthy Privacy ...} &&\cellcolor{yellowy!25} $\circ$&&\cellcolor{greeny!25} $\bullet$& &&&&& &&&&& &&&&& &&&&& &&& \\ \toprule
        \scriptsize{67. [Support] Selective ...} &&\cellcolor{yellowy!25} $\circ$&\cellcolor{greeny!25} $\bullet$&\cellcolor{greeny!25} $\bullet$& &&&&\cellcolor{yellowy!25} $\circ$&\cellcolor{yellowy!25} $\circ$ &\cellcolor{greeny!25} $\bullet$&&&& &&&&& &&&&& &&& \\ \toprule
        \scriptsize{68. Private Link} &&\cellcolor{yellowy!25} $\circ$&\cellcolor{greeny!25} $\bullet$&& &&&&&\cellcolor{yellowy!25} $\circ$ &&&&& &&&&& &&&&& &&& \\ \toprule
        \scriptsize{69. Anonymity Set} &&&&& &&&&\cellcolor{yellowy!25} $\circ$&\cellcolor{yellowy!25} $\circ$ &\cellcolor{greeny!25} $\bullet$&&&& &&&&& &&&&& &&& \\ \toprule
        \scriptsize{70. Active Broadcast of ...} &&\cellcolor{yellowy!25} $\circ$&\cellcolor{greeny!25} $\bullet$&&\cellcolor{greeny!25} $\bullet$ &\cellcolor{yellowy!25} $\circ$&&&&\cellcolor{yellowy!25} $\circ$ &&&&& &&&&& &&&&& &&& \\ \toprule
        \scriptsize{71. Unusual Activities}&&&&& &&&& \cellcolor{greeny!25} $\bullet$&\cellcolor{yellowy!25} $\circ$ &&&&& &&\cellcolor{greeny!25} $\bullet$&&& &&&&& &&& \\ \toprule
        \scriptsize{72. Identity Federation ...} &&&&& &&&&&\cellcolor{greeny!25} $\bullet$ &&&&& &&&&& &&&&& &&& \\ \toprule
        \scriptsize{73. Dynamic Location ...} &&\cellcolor{yellowy!25}$\circ$&&\cellcolor{greeny!25}$\bullet$& &&&&\cellcolor{yellowy!25}$\circ$&\cellcolor{yellowy!25}$\circ$ &\cellcolor{greeny!25}$\bullet$&&&& &&&&& &&&&& &&& \\ \toprule[0.32ex]
    
    \end{longtable}
\endgroup

In the initial move towards correlating the privacy patterns and the CPLF, it is useful to construct criteria showing the capability of the privacy patterns to apply various principles and rights of the CPLF. Table 6 aims at correlating the privacy patterns and the principles and rights of the CPLF, which will enable thorough and efficient compliance with the main principles and rights of the CPLF of the various data protection laws. Two criteria have been identified: the ($\bullet$) in the various cells determines a direct relationship, and the ($\circ$) shows an indirect relationship between the privacy patterns and the CPLF. Underlining the relationship between the privacy patterns and among the various principles and rights of privacy laws, the criteria indicates that most of the privacy patterns are associated with more than one principle or right, except the following patterns: \textbf{Discouraging Blanket Strategies}, \textbf{Reciprocity}, \textbf{Policy Matching Display}, \textbf{Incentivised Participation}, \textbf{Personal Data Store}, \textbf{Pay Back}, and \textbf{Negotiation of Privacy Policy}. This might be due to the nature of the privacy patterns as they are more specific than the principles or rights of the CPLF. Nevertheless, some of the principles and rights of the CPLF have not been associated with any of the privacy patterns, as seen in Table 6. These are the following principles: \textbf{Source}, \textbf{Cross-border Disclosure of Personal Data}, \textbf{Dealing with Unsolicited Data}, and \textbf{Adoption, Use or Disclosure of an identifier}. However, these principles of the CPLF are only limited to Australia's and New Zealand's data protection laws. With Regard to the rights of the CPLF, \textbf{Right to Data Portability}, \textbf{Right to Complain}, and \textbf{Right of Individuals not to  be Discriminated} are not achieved by any of the privacy patterns. This could be due to the nature of the law as being largely disconnected from the technical domain. Some of the provisions of laws, furthermore, require organisations to comply with the provisions rather than implementing them in the development phase. Therefore, it cannot be implemented in the technical domain.

\newpage
\bibliographystyle{unsrt}  
\bibliography{References}
\end{document}